\documentstyle[12pt,aaspp4]{article}
\begin{document}
\title{Some Revised Observational Constraints on the Formation and 
Evolution of the Galactic Disk}
\author{Bruce A. Twarog, Keith M. Ashman, and Barbara J. Anthony-Twarog}
\affil{Department of Physics and Astronomy, University of Kansas,
Lawrence, KS  66045-2151}
\affil{Electronic mail: twarog@kuphsx.phsx.ukans.edu,
ashman@kusmos.phsx.ukans.edu, anthony@kuphsx.phsx.ukans.edu}

\begin{abstract}
A set of 76 open clusters with abundances based upon DDO photometry
and/or moderate dispersion spectroscopy has been transformed to a common
metallicity scale and used to study the local structure and evolution
of the galactic disk. The metallicity distribution of clusters with 
galactocentric
distance is best described by two distinct zones. Between $R_{GC}$ = 6.5
and 10 kpc, the metallicity distribution 
has a mean [Fe/H] = 0.0 and a dispersion of 0.1 dex; 
there is, at best, weak evidence for a
shallow abundance gradient over this distance range. Beyond $R_{GC}$ =
10 kpc, the metallicity distribution has a dispersion
between 0.10 and 0.15 dex, but with a mean [Fe/H] = --0.3, implying a sharp
discontinuity at $R_{GC}$ = 10 kpc. After correcting for the discontinuity,
no evidence is found for a gradient perpendicular to the plane.
Adopting the clusters interior
to 10 kpc as a representative sample of the galactic disk over the last 7
Gyr, the cluster metallicity range is found to be approximately half
that of the field star distribution. When coupled with the discontinuity
in the galactocentric gradient, the discrepancy in the metallicity 
distribution is interpreted as an indication of significant diffusion of field
stars into the solar neighborhood from beyond 10 kpc. These
results imply that, contrary to earlier claims, the sun is {\bf not} 
atypical of the stars formed in the solar
circle 4.6 Gyr ago. It is suggested that the discontinuity
is a reflection of the edge of the initial galactic disk as defined
by the disk globular cluster system and the so-called thick disk; the
initial offset in [Fe/H] created by the differences in the chemical
history on either side of the discontinuity has been carried through
to the current stage of galactic evolution. If correct, diffusion coupled
with the absence of an abundance gradient could make the separation of
field stars on the basis of galactocentric origin difficult, if not impossible.
\end{abstract}
 
\section{Introduction}
As probes of galactic structure and evolution, open clusters
have proven to be valuable but limited. Their value lies in the
improvement in accuracy for distance determination, metal
content, and age produced by the collective stellar sample which shares
these properties. The limitations have arisen because of
their often sparse population of members, age restrictions imposed
primarily by tidal disruption, and the need to study an extended
region of the galaxy to obtain a statistically significant sample.
Despite these limitations, there has been increasing
interest in the cluster population, particularly
the older clusters, spurred on in recent
years by the availability of CCD cameras on telescopes of
modest size. The sample growth has led to a comparable growth
in the analysis of its composite properties as detailed in Janes \& Adler
\markcite{j8}(1982), Friel \markcite{f10}(1989), 
Janes \& Phelps \markcite{j9}(1994), Phelps {\it et al.} \markcite{p7}(1994),
Carraro \& Chiosi \markcite{c4}(1994), Friel \markcite{f11}(1995), and 
Scott {\it et al.} \markcite{s10}(1995), among others.

Despite the improvements, there are
two areas where our understanding of the cluster sample has changed
little in the last 15 years. Among open clusters there
is no evidence for an age-metallicity relation (AMR), though
a significant range in [Fe/H] exists among clusters older than 1 Gyr,
as first detailed in Hirshfeld {\it et al.} \markcite{h16}(1978). When combined with
the seminal study of the cluster galactic abundance gradient by Janes
\markcite{j5}(1979),
it is apparent that position within the galaxy plays a more critical
role in defining a cluster's properties than age. Though subsequent
estimates for the cluster abundance gradient have often been somewhat 
steeper than the original Janes \markcite{j5}(1979) value,
the basic result that the outer galactic disk is metal-poor relative to the
inner disk has been confirmed many times since, in particularly conclusive
fashion by Friel \& Janes \markcite{f13}(1993), supplemented by 
Thogersen {\it et al.} \markcite{t2}(1994; 
hereafter collectively referred to as FJ). 

More recently an even larger sample has been used by 
Piatti {\it et al.} \markcite{p8}(1995; hereinafter PCA) to derive a similar galactocentric abundance gradient
and, for the first time, an open cluster abundance gradient perpendicular
to the galactic plane, contradicting the conclusions of \markcite{f13}FJ. The unique 
nature of the latter result combined with a
revised calibration of the DDO system (Twarog \& Anthony-Twarog \markcite{t11}1996),
the photometric system adopted by \markcite{p8}PCA,
suggested the need for a closer look at the cluster sample and the reality
of the Z-gradient, where Z refers to the direction perpendicular to the
plane. We find
that the commonly accepted picture of a linear abundance gradient with
galactocentric distance ($R_{GC}$) coupled
with a large spread in abundance at a given age in the solar neighborhood
fails to account for the observations. Instead, the disk breaks up into
two distinct populations defined by a sharp boundary at $R_{GC}$ = 10 kpc.
The goals of this paper are to explain the changes in the cluster
analysis which are the foundation of the new sample (Sec. 2), to 
quantify the reality of the two cluster populations and the galactic
abundance gradient (Sec. 3), to compare and contrast the cluster
population with that of the field stars in the solar neighborhood,
and to attempt an explanation of the source of the apparent discrepancy
between the field stars and the clusters (Sec. 4). Our conclusions and
a qualitative scenario for the evolution of the disk are
summarized in Sec. 5.

\section{The Data}
Two key cluster parameters of interest are  metallicity and
the distance. Given these and galactic position, one can construct a
spatial map of the cluster population and study its global
characteristics. The number of clusters for which metallicity estimates
exist is well over a hundred, but such estimates involve an array of
techniques used by different observers with differing choices for cluster
membership, reddening, and abundance calibrations. 
Simple merger of all cluster abundances can wash out
or destroy detailed structure within the cluster population unless it
is done with extreme care. We will make use of only
two sources, the DDO sample described in \markcite{p8}PCA and the
spectroscopic sample in \markcite{f13}FJ. We have excluded the unpublished
abundances listed in Friel \markcite{f11}(1995) and the supplemental data
of Lyng\aa \markcite{l11} (1987) used by \markcite{f13}FJ and Friel \markcite{f11}(1995) because many of the
clusters contained in the supplemental sample have DDO data and, 
in key cases, the
parameters found in Lyng\aa \markcite{l11} (1987) have proven to be unreliable.

\subsection{Reddening and Distance Estimation}
The first step in the cluster compilation is the
determination of cluster reddening. No single technique for reddening
estimation has been applied to all the clusters in the sample, though the
DDO - $(B-V)$ technique of Janes \markcite{j4}(1977b) is applicable to a large fraction. 
We have surveyed the literature for each cluster, attempting
to assess the range and reliability of the published values. In the end
we have adopted a representative estimate for
each cluster. To make the most effective use of the reddening determinations
we have included an adjustment which is often neglected in cluster analyses.
It has been known since the work of Lindholm \markcite{l3}(1957) and 
Schmidt-Kaler \markcite{s9}(1961) 
that the degree of reddening
experienced by a star is dependent upon the color of the star. Generally,
a red giant exhibits a smaller $E(B-V)$ than a hotter main sequence star
when obscured by the same dust layer. Fernie \markcite{f5}(1963) has calculated that a
change in the intrinsic $(B-V)$ of 1.0 mag lowers the effective $E(B-V)$
by approximately 10\%. One of the few discussions and applications of this
phenomenon can be found in Hartwick \& McClure \markcite{h8}(1972). In deriving the
reddening appropriate for a cluster, we have taken into account whether
or not the estimate is based on the giants or hotter stars near the
turnoff. Final values are adopted individually for the main sequence and the
giant branch, fixed to ensure consistency based upon their relative colors.
The typical difference in $E(B-V)$ lies in the 5\% to
10\% range, leading to a minor difference for the majority of
clusters which have $E(B-V)$ below 0.10. Details for the 
individual clusters may be found in the
Appendix. In a majority of the cases our results differ
little from those listed in \markcite{p8}PCA and \markcite{f13}FJ, though controversy still surrounds
a number of key objects, e.g., NGC 6791. 

The grab-bag approach to reddening
may, at first glance, appear to violate the concern expressed above regarding
the merger of abundances. However, one of the strengths of both metallicity
techniques and one of the key reasons they were selected for this sample
is their weak sensitivity to reddening changes. For a change in $E(B-V)$ of 0.1
mag, the corresponding change in [Fe/H] is between 0.10 and 0.15 for
the photometric and the spectroscopic (Friel \markcite{f12}1997) approaches. Clusters for
which the uncertainty in the adopted reddening is larger than this have been
excluded from the discussion.     

Reddening values, though not critical to [Fe/H] because of
the weak sensitivity to changes in $E(B-V)$, can have a
significant impact on the cluster distance. However, the effect on 
distance depends on the technique used in obtaining the
distance. For the current investigation, three approaches have been used.

First, whenever adequate $BV$ photometry is available, distances 
have been derived via main sequence fitting.
As a tie-in to past work, the location of the main sequence for a cluster of
a given [Fe/H] is taken from the VandenBerg \markcite{v1}(1985) isochrones, adjusted
to guarantee that a solar mass star of age 4.6 Gyr with solar composition
has the adopted $(B-V)$ of 0.65 and $M_V$ = 4.84 (Twarog \& Anthony-Twarog
\markcite{t10}1989). Differential corrections to the distance due to differences between
the cluster [Fe/H] and that of the nearest isochrone set are applied using
the metallicity dependence as described in VandenBerg \& Poll \markcite{v2}(1989).

For the effect of reddening on distance determination via main sequence 
fitting, two factors compete. Increased reddening correction moves
the main sequence toward the blue. If the slope of the main sequence with
$(B-V)$ is X, the apparent modulus is increased by X$*E(B-V)$. However,
corrections to the distance modulus for increased
reddening make the true modulus smaller. The net change in the true 
modulus is (X -- 3.1)$*E(B-V)$. For the unevolved, cooler main sequence, X is
between 5 and 6. From the base of the turnoff, X rapidly increases, reaching
above 10 as the turnoff approaches vertical, or for the hotter main sequence
of younger open clusters.  Ideally, one would prefer to use the cooler,
unevolved main sequence to derive distances, but this is not always possible.
Main sequence fitting was applied to 49 clusters.

Second, for some older ($>$1.0 Gyr) clusters, the main sequence is too faint
or too ill-defined to permit reliable fits. If
it can be identified, however, the red giant clump provides an adequate
method of estimating the distance. Using 13 clusters with well-defined main
sequences and red giant clumps, we have derived the typical $M_V$ for the
clump using main sequence fitting. Assumed
reddenings and abundances are taken from Table 1. As illustrated in 
Fig. 1a, there is a weak dependence on metallicity, predominantly caused by the
most metal-rich cluster in the sample, NGC 6791. In Fig. 1b, the
dependence on age is investigated.  Rather than adopt an absolute age scale
and become involved in the controversy which often surrounds such choices, 
we have instead used an age ranking based upon the morphological
cmd parameter, MAR, as defined in Anthony-Twarog \& Twarog \markcite{a7}(1985) and revised
in Twarog \& Anthony-Twarog \markcite{t10}(1989). The MAR is a ratio of the magnitude
difference between the red giant clump and the brightest point at the
turnoff to the color difference between the red giant branch at the
level of the clump and the bluest point of the turnoff. Combining both
parameters enhances the age sensitivity while partially removing the
metallicity sensitivity of the cmd morphology. The ratio should also be
reddening-independent, but this assumption fails for clusters with
large $E(B-V)$ due to the reddening dependence on stellar color.
To allow readers
to place the trend in Fig. 1b on their preferred age scale, we have
tagged a few of the extreme clusters. The youngest objects are NGC 7789 and
NGC 3680, while the oldest are M67, MEL 66, NGC 188, BE 39, and NGC 6791.

Over the age range from NGC 7789 to MEL 66 (approximately 1 to 5 Gyr), 
the mean $M_V$ is 0.6 $\pm$ 0.1; the scatter is explained by the
uncertainties in the distance moduli and in the exact definition of the 
clump. We have attempted to choose the latter parameter based upon the
magnitude which is typical of the average star in the clump, rather than
the reddest or bluest point. From MEL 66 to NGC 6791 (approximately 5 Gyr
to 9 Gyr), the $M_V$ declines from 0.6 to 1.2 for NGC 6791. The fact that
NGC 6791 has both the largest age and the largest [Fe/H] makes it impossible
to decouple the effects of age and metallicity on $M_V$ among the oldest
clusters, though it does appear likely comparing Figs. 1a and 1b that age 
is the predominant effect in the increase in $M_V$. 

For most clusters in the 
age range of 1 Gyr to 5 Gyr, we will adopt $M_V$ = 0.6 for the clump. 
For metal-rich
and/or significantly older clusters, we will adopt a value between 0.7 and
1.0. The specifics for each cluster can be found in the Appendix. 
We note in closing that
Janes \& Phelps \markcite{j9}(1994) have attempted a similar estimate of the clump
luminosity, deriving $M_V$ = 0.90 $\pm$ 0.40 from 23 clusters older than
1 Gyr. The difference is easily explained. The cluster sample used by
Janes \& Phelps \markcite{j9}(1994) is a composite from a wide variety of sources.
Distance moduli depend upon reddening, metallicity, and the adopted
main sequence for the fit. The lack of normalization among the various
observers, particularly in isochrone fits from different theoretical
models, is a primary source of both the large scatter and the fainter
$M_V$. Second, the abundances for the clusters discussed in Figs. 1a and
1b are generally higher than found in past discussions of the clusters,
particularly those in the galactic anticenter. A larger [Fe/H] leads to
a larger modulus for a given main-sequence fit. The red giant clump was
used to derive distances for 4 clusters. 

For cluster distances derived by fitting the red giant clump to a fixed
$M_V$, the apparent modulus is fixed by the
apparent brightness of the clump. Changing $E(B-V)$ alters the true 
modulus by $A_V$ = 3.1$*E(B-V)$, making it smaller as $E(B-V)$ grows larger.
              
Third, if neither main sequence fitting nor a red giant clump match are
plausible means of deriving $(m-M)$, any technique available in the literature
is used. In most cases, this implies a distance determination tied to
a photometric $M_V$ calibration using intermediate or broad-band photometry
of the giants, as in DDO, or main sequence stars, as in $uvby$H$\beta$. 
The exact impact on the distance estimate of changing the reddening is
unique to each photometric system. This approach was adopted for 23
clusters.

As a prelude to the discussion in Sec. 3, we point out that estimation
of the uncertainties in the distance moduli is not trivial.
Even for a given approach, not all distances derived with that method
will have equal errors when using an amalgam of data sources. Though we have
taken the coward's way out and used the same uncertainty, $\pm$0.2, for all the
moduli, the conclusions of the investigation are only weakly dependent
upon uncertainties in the moduli. For main sequence fitting, in a
differential sense, the best cluster data are estimated to provide
moduli with an uncertainty between $\pm$0.1 and $\pm$0.2 mag; the same number
applies to the clump-based distances. For the weaker main sequence
data and the majority of the distances based upon other, primarily
photometric, techniques, the uncertainties are closer to $\pm$0.3
mag, and may be larger in extreme cases. However, because the uncertainty in 
the distance modulus produces a
percentage change in the absolute distance and the absolute error
projects into a smaller change in galactocentric distance, for clusters
within 2 kpc of the sun doubling the estimated uncertainty will have
little impact on our conclusions.

\subsection{Metallicity Estimation}
Ideally, one would prefer to discuss only abundances based upon one technique
and obtained by one observer. Since this is impossible we make use of two
samples, one photometric on a common system, DDO, and one spectroscopic
but observed by only one group. Even given only two techniques, it is
still important that the abundances from each are on a common scale.
The following sections will define the metallicity scale and the means
of standardizing the results.

\subsubsection{DDO Photometry}
The commonly adopted calibration of the $\delta$CN index of the 
DDO system for disk giants was 
first proposed by Janes \markcite{j2}(1975) based upon 44 stars. Since that time, 
suggested changes in the zero-point have been made by Deming {\it et al.}
\markcite{d4}(1977)
and Twarog \markcite{t8}(1981), while Luck \markcite{l10}(1991) revised both the slope and the zero-point.
Piatti {\it et al.} \markcite{p9}(1993) revised the definition of the CN-index, using an 
expanded sample of 82 G and K giants to redefine the DDO metallicity 
calibration in terms of $\Delta$CN, the technique applied to the open
clusters in \markcite{p8}PCA. Beyond the modest sample size, the primary shortcoming of
the earlier recalibrations tied to spectroscopic abundances is the
lack of consistency among the sources for the spectroscopic abundances.
This weakness was corrected in the comprehensive approach of
Taylor \markcite{t1}(1991), using over 300 field giants adjusted to a well-defined,
standardized system. Taylor \markcite{t1}(1991) concluded that the transformation
of the traditional $\delta$CN index to [Fe/H] required both a slope and
zero-point which were temperature-dependent.

The calibration adopted here is detailed in Twarog \& Anthony-Twarog
\markcite{t11}(1996).
Rather than use a composite
catalog of spectroscopic abundances, the DDO calibration was tied solely
to the data of McWilliam \markcite{m12}(1990), a sample of 671 field giants reduced and
analyzed in a uniform way. From 438 giants, Twarog \& Anthony-Twarog
\markcite{t11}(1996) 
confirmed the
color dependence of the calibration zero-point, but found a constant and
shallower slope than Janes \markcite{j2}(1975). It should be emphasized that 
the Taylor
\markcite{t1}(1991) and Twarog \& Anthony-Twarog \markcite{t11}(1996) calibrations, except for
a small zero-point shift to approximate a Hyades abundance of [Fe/H] =
+0.12, give very similar results. The use of a more internally consistent
spectroscopic sample does halve the scatter in the residuals attributable
to the DDO calibration, photometric errors aside, to less than
$\pm$0.05 for the bright star photometric sample of McClure \&
Forrester \markcite{m32}(1981). The revised 
calibration is applicable from [Fe/H] = +0.25 to --0.5. At lower 
abundances, the sample of McWilliam \markcite{m12}(1990) is inadequate to allow a
reliable calibration, but use of an expanded sample shows that the
calibration cannot be extrapolated. The $\delta$CN index increasingly
overestimates the abundance of more metal-deficient giants (Twarog
\& Anthony-Twarog \markcite{t11}1996).

For each cluster in the final sample of \markcite{p8}PCA, DDO photometry was collected
using the references cited by \markcite{p8}PCA, and expanded whenever possible by 
doing an updated search of the literature. An analogous approach was
applied to the reddening, radial-velocity data, and proper-motion
studies. Definite non-members were eliminated from the sample, but
stars with no membership information were included, as were stars
classed as binaries, unless otherwise noted in the Appendix. The
reason for this is that a large fraction of the clusters lack
definitive membership information and no binary classifications.
Inclusion of these stars in the mean metallicity estimate for
clusters with partial information places all the clusters on an equal
footing.  Moreover, it was found that exclusion
of binaries rarely had a significant impact upon the mean abundance
or the dispersion for a cluster.

Comparison of the results for individual clusters with those of \markcite{p8}PCA will
occasionally show differences in the number of stars included in the
cluster average. These differences are sometimes attributable to different
calibration limits for the two techniques, but often arise because
of differences in membership classification. Because \markcite{p8}PCA does not
provide specific details on which stars in each cluster are excluded,
we have no means of making an exact comparison. This problem is compounded
by the inclusion of unpublished radial-velocity data in deciding membership.

In a few extreme cases, the cluster abundances in \markcite{p8}PCA are based 
exclusively upon unpublished
photometry. Rather than drop these clusters from the sample, we have attempted
a simple transformation of the \markcite{p8}PCA abundances to our system in the following
manner. For clusters where the number of giants is large 
enough that small differences
in the total included have little impact on the cluster mean or where no
difference exists in the sample of giants included, the DDO metallicity
has been calculated using the calibration discussed above and compared
with that listed by \markcite{p8}PCA. To isolate the impact of the different calibrations,
for this comparison we have adopted the same $E(B-V)$ as \markcite{p8}PCA for each cluster.
The residuals in [Fe/H] in the sense (\markcite{p8}PCA -- TAT) are plotted in Fig. 2
as a function of the [Fe/H] in \markcite{p8}PCA (open circles). The 
scatter at a given [Fe/H] is
encouragingly small and is primarily caused by small differences in the
calibrations as a function of the color of the giant. For clusters with
[Fe/H] below --0.2 on the system of \markcite{p8}PCA, there appears to be a common
offset of about 0.1 dex between the two metallicity scales, with that
of \markcite{p8}PCA being more metal-poor. At higher metallicity, the residuals show
an approximately linear trend with increasing [Fe/H], implying that
the metallicity range among clusters on the scale of \markcite{p8}PCA is larger than
that found  on the revised scale. Based upon Fig. 2, we can transform
an abundances of \markcite{p8}PCA to the approximate scale of the revised
calibration using the following relations:

      [Fe/H]$_{\markcite{p8}PCA}$ $\leq$ --0.15 \ \ \ \ \ \  [Fe/H]$_{TAT}$ = [Fe/H]$_{\markcite{p8}PCA}$ + 0.09

      [Fe/H]$_{\markcite{p8}PCA}$ $>$ --0.15    \ \ \ \ \ \  [Fe/H]$_{TAT}$ = 0.55[Fe/H]$_{\markcite{p8}PCA}$ + 0.02

These transformations have been used only for clusters where the DDO
photometry remains unpublished. In all other cases the DDO calibration
of Twarog \& Anthony-Twarog \markcite{t11}(1996) has been applied individually to the giants.

\subsubsection{Spectroscopy}
Moderate-resolution spectra of cluster giants have been collected 
and calibrated 
by Friel and her coworkers over the last decade; the results are summarized 
in \markcite{f13}FJ.  The beauty of the approach 
is that in addition to metallicity, one
obtains an indication of membership via radial velocities of modest
accuracy. Though the emphasis of the sample is on older disk clusters,
a related byproduct is inclusion of the largest sample of abundances for
clusters at large galactocentric distances, the majority of which have
no DDO estimate due to the faintness of the giants. 

The current DDO sample has an overlap
of 14 clusters with \markcite{f13}FJ. As with the DDO sample, the literature on each
cluster has been reviewed to eliminate non-members and revise the reddening
estimate whenever necessary.  Because the techniques of \markcite{p8}PCA and \markcite{f13}FJ make use
of many of the same field giants in their calibrations, the abundances
of \markcite{p8}PCA and \markcite{f13}FJ agree well over the range in [Fe/H] except 
for a small zero-point shift. After adjusting the abundances for a common
$E(B-V)$ and a shift of +0.05 in [Fe/H] to place them on the same system as
\markcite{p8}PCA, one derives the residuals in [Fe/H] in the sense (\markcite{f13}FJ -- TAT), plotted
in Fig. 2 as crosses. Within the uncertainties, the residuals
follow the trend consistent with the transformation defined by the data
of \markcite{p8}PCA; the mean [Fe/H] estimate for each cluster of \markcite{f13}FJ has been 
adjusted in this manner. For clusters common to the two
samples, the transformation based upon the cluster mean has been 
applied individually to each giant in \markcite{f13}FJ, and the results for
the spectroscopic and photometric approaches merged to define the
mean and the standard deviation used in the final analysis for the cluster.

It should be emphasized that the inclusion of the spectroscopic sample removes
one potential source of bias from the analysis. Because of the insensitivity
of DDO below [Fe/H] = --0.5, it is possible that an apparent 
metallicity cutoff could occur in a purely DDO-based sample, shifting the
mean of the abundance distribution and artificially decreasing the dispersion.
For every cluster except one with [Fe/H] below --0.3, the abundance is
based upon either the spectroscopic data or the combined DDO and spectroscopic
abundances. As discussed above and illustrated in Fig. 2, the offset between
the DDO and the spectroscopic scales at lower metallicity is constant and
well-determined; it exhibits no evidence for changing sensitivity over the
range of interest. NGC 2204 at [Fe/H] = --0.34, well above the sensitivity
cutoff, is the only metal-poor cluster with a metallicity from DDO data
alone.

\subsection{The Cluster Parameters}
The abundance results for the clusters are summarized in Table 1. Columns
1 through 3 give the cluster identification and its galactic coordinates.
Columns 4 and 5 list the adopted reddening $E(B-V)$ as defined by the stars
at the cluster turnoff and by the red giants, respectively. Column 6 explains
the source of the metallicity estimate: DDO means [Fe/H] from the red giants
based upon the calibration of Twarog \& Anthony-Twarog \markcite{t11}(1996); DDT means
the abundance of \markcite{p8}PCA adjusted for reddening and transformed using the
relations discussed above; and SPE implies the spectroscopic abundances of
\markcite{f13}FJ, adjusted for reddening, shifted by +0.05 in [Fe/H], and transformed
using the relations cited above. For the 14 clusters that have both DDO and
SPE abundances, an additional line is included and tagged with DSP. For this
line, the abundance listed is the average abundance using
the DDO sample and the SPE sample together. In the discussions which follow, 
this is the source of the [Fe/H] adopted for these clusters.
The [Fe/H] is presented
in column 7, followed by the standard deviation of the sample, the number
of stars included in the average, and the standard error of the mean for
[Fe/H].  For clusters with only two giants, the standard deviation and the
standard error of the mean have been set to one-half the difference in
[Fe/H] between the stars. For clusters with only one giant, the errors
quoted are the average standard deviation in [Fe/H] for a single star
based upon clusters with DDO abundances from 3 or more giants.

In Table 2 one can find the information relating to the galactic
properties of the cluster system. Columns 1, 2, and 3 give the identification,
the mean [Fe/H], and the standard error of the mean from Table 1, respectively.
Column 4 identifies the
means of estimating the distance modulus: MSF is main sequence fitting with
the reddening based primarily upon the turnoff stars; RGC is the assumed
red giant clump with the reddening based primarily upon the giants; and
OTH is whatever additional method is available. For details, the reader is
referred to the Appendix. Column 5 lists the apparent modulus, followed
in columns 6 and 7 by the adopted apparent magnitude of the clump, if
it can be identified, and $M_V$ for the clump, derived if MSF is listed but
assumed if RGC is present. Columns 8, 9, and 10 list the distances in kpc
from the sun, the galactic center, and the galactic plane, respectively,
on a scale where $R_{GC}$ for the sun is 8.5 kpc.
   
\section{The Cluster Abundance Pattern with Position}
\subsection{The Galactocentric Abundance Gradient}
The data in Table 2 are plotted in Fig. 3a with error bars and
in Fig. 3b without error bars for clarity. In Fig. 3b open circles
represent rederived abundances from DDO data of indivdual stars, squares are
DDO cluster abundances transformed directly from the system of \markcite{p8}PCA to
the revised scale, open triangles are clusters with abundances from
spectroscopic data, and filled triangles are abundances from combined
spectroscopic and photometric methods. The two points joined by a line are the
limiting values for BE 21 as discussed in the Appendix. Errors in 
[Fe/H] are the
standard errors of the mean. The error bars in $R_{GC}$ have been derived
assuming that the uncertainty in the apparent modulus is the same for
all clusters, $\pm$0.2 mag. Though on a relative scale this is probably
an overestimate for clusters whose distances are based upon either main
sequence fitting or the $M_V$ of the red giant clump, it is likely to be
an underestimate for the remaining clusters. Again, for specific clusters
the reader is referred to the Appendix. The uncertainty in $R_{GC}$ is derived 
by determining the galactocentric distance over the range in $(m-M)$.
Clearly, the closer the cluster is to the sun, the smaller the absolute
error in the galactocentric distance. Moreover, the error in $R_{GC}$ is
minimized for clusters with galactic longitude near 90$\arcdeg$ and 270$\arcdeg$.
Thus, the errors increase with increasing distance from the sun 
for clusters in the direction of the galactic center and anticenter.

It is clear from Fig. 3 that a significant change in metallicity does
occur over the galactocentric range from 6.5 kpc to 15 kpc. What is not
clear is the validity of the assumption that the change is linear.
Not only do the clusters below [Fe/H] = --0.2 lie preferentially in
the galactic anticenter, they are located exclusively beyond $R_{GC}$ =
10 kpc, while not a single cluster with [Fe/H] $>$ --0.15 occupies the
same region. Rather than a linear transition with distance, it appears
that the galactic disk contains a relatively abrupt discontinuity near
10 kpc. 

To begin the evaluation of the significance of this feature, we first
attempt the traditional approach of simply fitting a line through the data. 
We have derived the least-squares fit under three
circumstances: exclusion of BE 21, inclusion of BE 21 with the low 
[Fe/H], and inclusion of BE 21 with the high [Fe/H]. The results are
summarized in Table 3.

In the first three cases, the slope of the abundance gradient ranges
from --0.077 kpc$^{-1}$, if the low [Fe/H]
data for BE 21 are used, to --0.067 kpc$^{-1}$ if BE 21 is excluded completely.
In all three cases, the uncertainty in the slope is $\pm$0.008, smaller
than any previous cluster study of the abundance gradient. Moreover,
the probability that a correlation coefficient near 0.75 can arise
from a truly random sample of this size ($P_R$) is well below one part in ten 
thousand.

What happens if we artificially break the sample into two groups based upon
their location within (62 clusters) or beyond (14 clusters) $R_{GC}$ = 10 kpc?  
For the inner group, a small gradient persists, but its statistical
significance is marginal; the correlation coefficient is reduced to 0.22,
which has a 9\% probability of coming from a purely random sample. If we
weight the data using the inverse of the standard error of the mean in 
[Fe/H] to enhance the impact of the clusters with more reliable abundances,
the gradient weakens in size and statistical significance; there is a 
19\% probability that the derived gradient of --0.023 kpc$^{-1}$ comes
from a random sample. For the outer group, the small sample size is
a severe limitation, though the range in $R_{GC}$ is greater than that
for the inner group. Of the three possible cases, only the one including
BE 21 with the low [Fe/H] parameters comes close to producing a significant
gradient and even this value is marginal. We point out that while we have
excluded the unpublished spectroscopic abundances of Friel \markcite{f11}(1995) because
they are preliminary in nature and lack fundamental details such as 
reddening estimates and error bars, the starred symbols in Fig. 7 
of Friel \markcite{f11}(1995) are in excellent agreement with the above interpretation 
and would only enhance the result if included. Only the most distant
cluster, BE 20, falls just outside the metallicity range defined by
the clusters between 10 and 13 kpc.

Our interpretation of these results is simple. The dramatic change that
occurs in the statistical significance of the gradient when one shifts
from the complete sample to two subgroups divided purely on the basis
of galactocentric position implies that rather than a continuous decline
in [Fe/H] between $R_{GC}$ = 6.5 kpc and 15 kpc, there are actually
two distinct groups of clusters. Within each group, the metallicity 
gradient is weak to nonexistent. The primary
difference between the two groups is in their mean metallicity; the outer
clusters are, on average, 0.3 dex more metal-poor than the inner
clusters. Thus, the only characteristic of significance in determining
group membership for a cluster is galactocentric position.

Unfortunately, one could argue that the sample breakpoint used in the
analysis is hardly objective; it was chosen specifically because Fig. 3
indicated that the gradients on either side of the breakpoint were weak.
To approach the analysis
somewhat more objectively, we have employed the KMM mixture-model algorithm
(McLachlan \& Basford \markcite{m10}1988; Ashman {\it et al.} \markcite{a14}1994) 
which can explore the presence and significance of multiple
peaks within the cluster metallicity distribution. It should be noted
that the sample is neither random nor complete; all clusters of all ages
have not been included at all galactocentric distances. There are clearly
fewer clusters at large galactocentric distances and the majority of
these are older than 1 Gyr, in contrast with the inner clusters which include
a large fraction of younger objects. However, while the sample is not ideal,
it is not biased on the basis of metallicity, i.e., the clusters that
were selected for photometric and/or spectroscopic analysis were not
chosen because they had peculiar or extreme abundances. If they are
representative of the clusters at their galactocentric distance, then
any structure found within the metallicity distribution must be tied to
correlated structure in the galactocentric distribution. 

The KMM algorithm objectively partitions a dataset into statistically-preferred
groups and quantifies the improvement in the fit relative to a single group.
These groups are fit by Gaussians either with the same variance (homoscedastic
case) or unequal variance (heteroscedastic case). 
The algorithm returns a probability, $P_{KMM}$, which is a measure of the improvement
of a multi-group fit to the data over a single Gaussian. $P_{KMM}$ values
below 0.05 represent significant rejections of a single Gaussian, while
values in the 0.05 to 0.10 range indicate marginal rejection of the
single Gaussian hypothesis. Full details are provided by Ashman {\it et al.} 
\markcite{a14}(1994).

In the present case, we have used the algorithm to
determine whether two Gaussians provide a better fit to the cluster
metallicity distribution than a single Gaussian.  Table 4
gives a summary of the probabilities, means, dispersions, and the numbers
of clusters in each group for the four options attempted: homoscedastic
and heteroscedastic, low-[Fe/H] BE 21, high-[Fe/H] BE 21. 
The derived $P_{KMM}$ indicate that in all cases a single Gaussian fit can be
rejected; significant improvement always occurs with the adoption of
two groups. For the low-[Fe/H], homoscedastic case, the cluster sample
divides into two groups with only 8 clusters in the low-metallicity
camp, just over half the outer clusters. The small number of metal-poor
clusters is a product of the low [Fe/H] for BE 21 and the 
constraint that the two groups have identical variances. In the less
restrictive heteroscedastic case,
two groups are again found, but now 19 clusters populate the metal-poor
sample, ranging from [Fe/H] = --0.15 to --0.83. The breakdown of the
sample supports the discontinuity discussed previously. None of the
outer clusters is assigned to the high-metallicity camp. The five clusters
with R$_{GC}$ $<$ 10 kpc assigned to the low metallicity group have
[Fe/H] in the range from --0.15 to --0.18 and represent the five most
metal-rich clusters in the low metallicity bin.

If one adopts the high-[Fe/H] value for BE 21, the partition of the
sample into two groups remains almost unchanged for the heteroscedastic
case. The variance of the low metallicity group increases and an additional
inner cluster with [Fe/H] = --0.13 is included. The means are essentially
unaltered. For the homoscedastic case, the separation by position is even
more apparent. The low metallicity bin contains 12 clusters, all beyond
10 kpc and in the [Fe/H] range from --0.54 to --0.24; only two outer
clusters near [Fe/H] = --0.2 are classed in the metal-rich category.
Given the uncertainty in [Fe/H] for many of the clusters under discussion,
the fact that the partitions in metallicity correspond closely to our
division of the clusters about the 
critical location at $R_{GC}$ = 10 kpc is an impressive confirmation
of what we believe is apparent in Fig. 3.

To emphasize the reality of the discontinuity,
we have binned the cluster sample into five groups purely on the basis 
of $R_{GC}$. As listed in Table 5, the spacing has been selected to
provide similar numbers of clusters in each bin, rather than identical
spacing in galactocentric distance, with the result that all the clusters
beyond $R_{GC}$ = 10 kpc fall within one bin. We emphasize that the
spacing of the bins for clusters interior to 10 kpc is irrelevant; all
that matters is the positioning of the last bin beyond 10 kpc. The abundance
distribution of each bin has been analyzed with the ROSTAT package (Beers
{\it et al.} \markcite{b2}1990; Bird \& Beers \markcite{b4}1993) 
to test its consistency with
a Gaussian; except for the innermost bin, all distributions
are consistent with a Gaussian.
The metallicity distribution of the inner bin is skew and marginally
inconsistent with a Gaussian ($P$ = 0.070).

For each of the metallicity distributions we have calculated the biweight
estimators of location ($C_{BI}$) and scale ($S_{BI}$). 
These are robust estimators analagous to the
familiar classical mean and dispersion of a distribution. These estimators
are less sensitive to outliers than their classical counterparts---a
property which can be particularly useful when dealing with samples
with a small number of points (see Beers {\it et al.} \markcite{b2}1990 and 
Bird \& Beers \markcite{b4}1993 for a full discussion). Our conclusions 
are unaltered if we use the classical mean and dispersion in
this analysis. Also included in Table 5 are 90\% confidence
intervals on these parameters calculated with a bootstrap resampling
technique included in the ROSTAT package. 
 
It is apparent that inside 10 kpc, there is no evidence for
a significant gradient. The mean metallicities are consistent
within the errors. This changes radically for the last bin, where the
mean metallicity drops by 0.30 to 0.35 dex. (Note that use of either 
[Fe/H] for BE 21 gives a similar result for the bin mean. This 
apparently contradictory result stems from the use of the biweight
estimators which gives lower weight to the more extreme data for
BE 21 in the low-[Fe/H] case.) What
is also intriguing is the observation that the scale of the [Fe/H]
distribution
is effectively constant near $\pm$0.10 for all bins.
The metallicity 
dispersion is important because it includes the scatter caused by
the abundance determinations, by any real galactocentric abundance gradient,
by any age spread, and finally by any intrinsic spread in [Fe/H] at a given
position at the time of formation of the clusters.

A final consistency check that supports our interpretation of a metallicity
discontinuity is provided by correcting the sample 
for the best-fit single linear
gradient and the two weak gradients found when the clusters are divided
into the two groups interior and exterior to $R_{GC}$ = 10 kpc. Using the 
gradients listed in Table 3, we correct the metallicity of each cluster
based on its galactocentric distance. Such
a correction removes the contribution to the dispersion of the metallicity
distribution produced simply by either a gradient or a discontinuity. Further,
one expects that the better description of the data will lead to a lower
dispersion of the resulting corrected metallicity distribution. 

The uncorrected scale of the total cluster sample is
$S_{BI} = 0.156$ (0.128, 0.186) dex, where values in brackets represent the 90\%
confidence limits on this quantity. (We have used the high [Fe/H] value
for BE 21: using the low value leads to similar results, as does
the use of the classical dispersion
rather than $S_{BI}$.) The scale
of the distribution when corrected for a single linear gradient is
0.113 (0.100,0.126), whereas the scale after a correction based on
the two discontinuous weak
gradients is 0.097 (0.087,0.108). While the 90\% confidence limits on these
two values overlap, this is suggestive that a discontinuity in metallicity
with galactocentric distance is a better interpretation of the cluster data.
If this is the case, we predict that an expanded cluster sample will
definitively show a lower dispersion for the metallicity distribution 
corrected for this discontinuity than a correction for a single
linear trend. The metallicity spread
among the clusters in the solar neighborhood will be the focus of Sec. 4.

In light of the sharp contrast between the commonly accepted view of a linear
abundance gradient and that of a discontinuous disk, one might ask why
this feature has remained hidden for so long. The straightforward answer is 
that the feature has been noted in the past (e.g., Janes \markcite{j5}1979; 
Panagia \& Tosi \markcite{p1}1981; \markcite{f13}FJ; Friel \markcite{f11}1995). However, 
the previous samples 
were inadequate to guarantee the reality of structure at this level or
the feature was interpreted as a steepening of the gradient with distance,
rather than a discontinuity. A recent example of this effect  
is seen in the work by Moll\'{a} {\it et al.} \markcite{m27}(1997) using a composite sample of
clusters from Panagia \& Tosi \markcite{p1}(1981), Cameron \markcite{c1}(1985), and 
Friel \markcite{f11}(1995).
The overall gradients at different ages for the open clusters range from
--0.089 $\pm$ 0.025 for young objects, to --0.072 $\pm$ 0.020 for
intermediate age, to --0.115 $\pm$0.037 for old clusters. What is 
intriguing is the result for the intermediate age group when divided
into two samples at $R_{GC}$ = 9 kpc. The inner clusters have a gradient
consistent with zero, while the outer clusters
show a gradient of --0.083$\pm$0.027. The presence of the outer gradient
reflects the inclusion of clusters between 9 and 10 kpc in the analysis
of the outer sample. 
The cluster sample in Tables 1 and 2
supercedes all past analyses in terms of the number of clusters, the
range in galactocentric distance, and the
internal consistency of the metallicity and distance scales. With
significant scatter among the data points, it is plausible that a step 
function could
be smeared into a linear gradient; the reverse process seems highly
implausible.

Is it possible that the discontinuity is a product of the recalibration 
of the DDO metallicity scale or of the cluster selection? Though the
revised DDO scale has compressed the range, primarily at the metal-rich
end, it has not changed the metallicity ranking of the clusters.
As noted earlier, the possibility of a bias due to the decline in
DDO sensitivity at lower [Fe/H] is removed through the inclusion of
the spectroscopic data. Direct
use of the \markcite{p8}PCA or \markcite{f13}FJ scale will not remove the discontinuity. The outer
disk clusters would all be systematically lower in [Fe/H] by about 0.1 dex;
the scatter among the inner clusters will broaden while the mean
[Fe/H] shifts down by about 0.05 dex. The discontinuity is actually
enhanced.

As for cluster selection, there is one source of concern. The cluster
data of \markcite{f13}FJ, which dominates the analysis in the anticenter, is
predominantly composed of clusters older than 1 Gyr. The sample inside
$R_{GC}$ = 10 kpc is heavily weighted toward younger open clusters.
If there is a significant AMR among clusters, the
discontinuity might be a reflection of the relative contributions
of young and old clusters in different galactocentric regions. Though the
sample is not as large as one might wish, if the clusters with ages
beyond 1 Gyr alone are analyzed, this concern evaporates. There is little
evidence for a cluster AMR among either the inner clusters or the outer
clusters alone; the range in age covered by the two groups is the same,
though the outer cluster sample is larger. Strangely enough,
the oldest clusters in the two groups (NGC 6791 and BE 39)
are also among the most metal-rich in each group. 

The age
distribution might play some role in the apparent increase in the mean
[Fe/H] for the innermost bin in Table 5. van den Bergh \& McClure \markcite{v5}(1980)
and Janes \& Phelps \markcite{j9}(1994) have pointed out the sharp drop in the number of older
clusters interior to $R_{GC}$ = 7.5 kpc. Though a large increase in
[Fe/H] with decreasing age is excluded by the observations, a change
of 0.05 to 0.10 dex over the last 5 Gyrs (Twarog \markcite{t7}1980b; 
Meusinger {\it et al.}
\markcite{m21}1991) is well within the errors. If the clusters in the innermost
bin are exclusively less than 1 Gyr in age, the mean [Fe/H] should be
higher, independent of any radial abundance gradient.

Aside from clusters, is there any evidence to support the notion of
an abundance discontinuity in the galactic disk? For reasons which
will be discussed more fully in Sec. 4, we exclude the studies based upon field
stars unless the stars are recently formed. Unlike clusters, field
stars have the ability to diffuse over large distances on Gyr
timescales (Wielen \markcite{w3}1977), distorting if not destroying potential
fine structure in the disk. Young stars, however, have not had enough
time to move significant distances from their place of origin and should
provide a reliable indicator of the current disk gradient. The best way
to guarantee a sample of truly young stars is to pick those of high 
mass, though this can lead to difficulty in estimating metallicity and 
in comparing it with values derived from different techniques for cooler
giants or lower mass dwarfs. Two stellar samples 
in the recent literature are of particular relevance, the spectroscopic 
analyses of Cepheids by
Fry \& Carney \markcite{f15}(1997) and of B stars in young clusters and associations
by Smartt \& Rolleston \markcite{s14}(1997).

Fry \& Carney \markcite{f15}(1997) obtained high dispersion spectroscopy of 23 Cepheids
over $R_{GC}$ = 6 to 10 kpc; 18 of the stars were observed at more than
one pulsational phase to test for any [Fe/H] dependence on the temperature
scale and dwarfs were observed in two clusters to test for non-LTE
effects. The mean metallicity of the sample is [Fe/H] = --0.05, which
probably implies a small zero-point shift relative to our scale. What
is more important is that over this galactocentric distance range, if
one Cepheid at $R_{GC}$ = 7.5 kpc with an anomalously low [Fe/H]
is excluded, the derived abundance gradient is --0.003 $\pm$ 0.018
with a correlation coefficient of only 0.04. Though Fry \& Carney
\markcite{f15}(1997)
justifiably caution against accepting this result as statistically
significant given the modest galactocentric distance range, the
lack of a measurable gradient, as well
as the modest dispersion in [Fe/H] at a given $R_{GC}$, are
clearly consistent with the cluster data over the same region.

Smartt \& Rolleston \markcite{s14}(1997) have compiled and analyzed in a homogeneous
manner spectra of 21 B stars in open clusters and the field. Their
metallicity indicator is [O/H], not [Fe/H], so some concern exists
over the relevance to the current discussion. Since the nucleosynthetic
origins of O and Fe are different, it is possible for a galactic
gradient to occur in one element and not the other. The focus of the
B star analysis is the apparent discrepancy between gradients identified
through [O/H] measures in nebular regions (HII and planetary nebulae)
and B stars. Abundance gradients of d[O/H]/dR = --0.06 $\pm$ 0.02 kpc$^{-1}$
are found in all the nebular studies, while B stars exhibit little or no 
gradient. Since B stars should be representative of the current
interstellar medium, this presents a problem. Smartt \& Rolleston
\markcite{s14}(1997)
attribute the lack of a gradient in previous studies to small samples,
and errors in distances and abundances; they derive a gradient of --0.07
$\pm$0.01 kpc$^{-1}$. Closer examination of Fig. 1b of Smartt \& Rolleston
\markcite{s14}(1997) shows that an alternate interpretation is well within the
errors of the data. The 11 B stars between $R_{GC}$ = 6 and 10 kpc,
the majority of which have small uncertainties in both distance and
abundance, exhibit no gradient at all. The mean [O/H] is --0.06 with a
dispersion of only 0.09 dex, fortuitously similar to the Cepheid result.
Again, we caution about overinterpreting the absolute abundances given
the possibility of scale shifts, but the modest dispersion is real.
Given that the errors in the [O/H] determinations are comparable to
the dispersion, this implies that all the B stars have the same abundance,
within the uncertainties. The entire source of the gradient comes from
the 10 points beyond 10 kpc, the majority of which have significantly
larger errors in both distance and abundance. There is no question
that, in the mean, [O/H] of the outer disk stars is lower by about
0.4 dex; whether this is due to a gradual change in [O/H] with distance or
a discontinuity remains determined by the eye of the beholder.  
In contrast, the lack of a gradient between $R_{GC}$ = 6 and 10 kpc from these
very young stars is in direct contradiction with the results of
Luck \markcite{l9}(1982), who finds d[Fe/H]/dR = --0.13 $\pm$0.03 kpc$^{-1}$
from 50 late type supergiants over $R_{GC}$ = 7.7 to 10.2 kpc.

An equally tantalizing picture is painted by 
the nebular results.
Most recent work (e.g., Fich \& Silkey \markcite{f6}1991; 
Maciel \& Koppen \markcite{m1}1994; 
Simpson {\it et al.} \markcite{s13}1995) indicates a variation in the abundances
of a number of elements across the galactic disk, confirming the earlier
work by Shaver {\it et al.} \markcite{s11}(1983). However, the number of points, the
analytical approach, and the range
of galactocentric distance varies significantly from study to study, and
none of the studies measure Fe. Thus, one is 
faced with many of the same problems that
plagued past attempts at deriving fine structure within the disk using
stars and clusters. It is intriguing to find, however, that despite the
problems, possible evidence for fine structure is not absent from the nebular
surveys. Simpson {\it et al.} \markcite{s13}(1995) claim that their abundance data, ranging
from $R_{GC}$ = 0 to 10.5 kpc, can be described by a linear gradient with
the mean abundance decreasing with distance
or equally well by two zones without gradients but linked via 
a discontinuity in the abundances  at 6 kpc, i.e.,
there is no abundance gradient between 6 and 10 kpc from the galactic
center. Though their data end where our discontinuity begins, they
point to the work of Fich \& Silkey \markcite{f6}(1991), Dinerstein {\it et
al.} \markcite{d5}(1993),
and still unpublished results to
suggest that a second discontinuity does exist beyond 10 kpc. 

The conclusion that a step function fits the data interior to 10 kpc as
well as a linear relation has been challenged by Afflerbach {\it et al.}
\markcite{a2}(1997)
using 34 compact HII regions between 0 and 12 kpc. They find linear gradients
of approximately --0.07 kpc$^{-1}$ for [N/H], [S/H], and [O/H], uncertainties
in the slopes comparable to our linear fits for the entire cluster sample,
and correlation coefficients near 0.7. They exclude a step function because
analysis of only HII regions beyond 6 kpc does not eliminate the gradients,
though they are reduced. However, of the 18 regions beyond 6 kpc, 4 lie
beyond the discontinuity at $R_{GC}$ = 10 kpc. The key question is
not if the data can be fit by a linear relation; they can be. What
matters is whether or not one can, given the sample size and the error
bars, distinguish between a step function and a linear relation. The
nebular data to date are, at best, inconclusive. Additional
concern comes from the absolute abundances derived from the nebular
samples, a point we will return to in Sec. 4.   
 
\subsection{The Abundance Gradient Perpendicular to the Plane}
A prime motivation for this investigation was the claim by \markcite{p8}PCA that a gradient
in [Fe/H] existed among clusters perpendicular to the plane, contradicting
earlier cluster analyses but consistent with field star studies (e.g.,
Yoss {\it et al.} \markcite{y1}1987; Sandage \& Fouts \markcite{s2}1987; Yoshii
{\it et al.} \markcite{y2}1989; Ratnatunga \& Freeman \markcite{r2}1989; 
Morrison {\it et al.} \markcite{m30}1990).
With the recognition that the cluster distribution is approximately
a step function rather than a linear gradient, it is straightforward
to show that the gradient perpendicular to the plane is an artifact
of the \markcite{p8}PCA analysis. In Fig. 4, the absolute Z distance is plotted for
all the clusters in Table 3 as a function of [Fe/H]; open circles are
clusters included in \markcite{p8}PCA while crosses are clusters added to the
current sample primarily through the data of \markcite{f13}FJ. No correction has
been applied for the galactocentric trend and a Z-gradient is obvious. 
\markcite{p8}PCA next applied a linear correction
to the sample to eliminate the radial trend and found that a residual
gradient still remained away from the plane. The source of the problem
is seen in Fig. 5, where the absolute Z position is presented as a
function of $R_{GC}$. Though there are some older clusters within
$R_{GC}$ = 10 kpc which are located well away from the plane, the
majority of the clusters are younger and lie within 200 pc of the
disk. In contrast, the clusters beyond 10 kpc are predominantly older
and are positioned well away from the plane. Thus, the
discontinuity in [Fe/H] is paired with a discontinuity in Z distribution.
For purposes of resolving the question at hand, it makes no difference
whether this  change is real or simply a selection effect in the cluster
sample. Note also that the cluster sample of \markcite{p8}PCA exends just beyond
the discontinuity. If, instead of correcting for a linear gradient,
one merely applies a zero-point offset of 0.32 dex to the outer clusters,
one gets the revised version of Fig. 4 as presented in Fig. 6;
the Z-gradient disappears.

\section{The Metallicity Distribution in the Solar Neighborhood}
\subsection{Galactic Clusters and Field Stars: The Discrepancy} 
Questions of the galactocentric gradient aside, the cluster sample
allows one to place another constraint on the chemical history of the
galactic disk. Assuming that the clusters are not affected by a significant
AMR, that clusters are not atypical of the interstellar medium in the
disk at the time of their formation, and that the intrinsic metallicity
spread within the interstellar medium does not change significantly
over the lifetime of the disk, one can use the metallicity distribution
among the clusters to sample the degree of inhomogeneity among 
the stars forming at a random time within the disk. For our sample, we
will only use clusters between $R_{GC}$ = 6 and 10 kpc. A small correction
to each [Fe/H] based upon galactocentric position and the derived
small abundance
gradient among the inner clusters (see Table 3, unweighted) has been applied. 
For our purposes it
is irrelevant whether this gradient exists because of an intrinsic
gradient with position at a given age, because of an AMR convolved
with a change in the mean age of the sample with galactocentric distance,
or both. Our only interest is in the dispersion in [Fe/H] at a given
location at a given time. From 62 clusters, the corrected data have 
a mean [Fe/H] of +0.010 and a robust dispersion of only
$\pm$0.096 (0.085, 0.108). The distribution is slightly boxier than a
Gaussian. Because this dispersion includes any residual trends with
age or galactocentric position and the observational uncertainties in the 
abundance estimates,
it should be regarded as an upper limit to the intrinsic metallicity dispersion within
the interstellar medium.

This dispersion seems small compared to past discussions of
the cluster distribution (e.g., Carraro \& Chiosi \markcite{c4}1994; Friel
\markcite{f11}1995), but it 
should be remembered that we have
excluded the clusters beyond 10 kpc, the abundance correction due to
the gradient is based solely upon the inner clusters and therefore actually
narrows the dispersion, and the change in the metallicity scale for the
giants has compressed the previous scale for the inner clusters. 
That is not to say that a 
true range in [Fe/H]
does not exist among clusters formed at approximately the same time; it does.
This result merely corroborates the earlier investigations of smaller
samples with precise abundances by Nissen \markcite{n1}(1988) and Boesgaard \markcite{b5}(1989).
Nissen \markcite{n1}(1988) used $uvby$H$\beta$ photometry of F-dwarfs in 13 nearby
open clusters with ages between 0 and 2 Gyr to study the metallicity
spread. The mean [Fe/H] = +0.05 with a dispersion among the clusters
of only $\pm$ 0.08; no correlation was found between [Fe/H] and age.
For reference, 9 of the clusters in Nissen \markcite{n1}(1988) are found in the
current investigation; the mean difference in [Fe/H] in the sense
(NI -- Table 2) is 0.00 $\pm$ 0.10.

Boesgaard \markcite{b5}(1989) used high dispersion spectroscopic analysis of F dwarfs
in 6 galactic clusters, all younger than 1 Gyr, to study the metallicity
spread in the solar neighborhood. Though the sample is small, the 
abundance estimates have unusually high precision for open clusters.
Boesgaard \markcite{b5}(1989) finds a mean [Fe/H] of +0.015 and a dispersion of only
0.087. These two studies are consistent with what is found above: the
[Fe/H] range among open clusters in the solar neighborhood is
between 0.3 an 0.4 dex wide, from
[Fe/H] = --0.2 to +0.2, with no evidence that this has changed
significantly over the last 5 Gyr, and a mean near solar. 
(A note of clarification: the 
parameter used to describe the inhomogeneity in [Fe/H] among stars
and clusters in the solar neighborhood varies from study to study.
The distribution of clusters inside 10 kpc is robust fit to a Gaussian though
the distribution is not a perfect match to a Gaussian. The dispersions quoted
for the smaller studies do not assume a Gaussian profile and represent
the traditional standard deviation about the mean. Our conclusions are
unchanged if we use the more classical measure of the dispersion. The range in
[Fe/H] is more appropriate when a Gaussian distribution is a poor 
representation, e.g., when the sample of all the clusters is best represented
by the sum of two Gaussians. As observed above, the range is about two to
three times larger than the dispersion, a fact which should be kept in
mind when comparing the results from different investigators.)
 
How does this compare with the results from field stars within the
solar neighborhood? In Sec. 3, we discussed the results for young stars
as defined by the Cepheids and the B stars inside 10 kpc. In both 
instances, the dispersion in metallicity from either O or Fe is typically
$\pm$0.10 dex or less with a mean abundance near solar, in excellent 
agreement with the cluster data. Venn \markcite{v10}(1995) derives a mean
of solar abundance for 13 metals from spectroscopic analysis of 22 A 
supergiants, though the abundance dispersion is closer to $\pm$0.2 due
to the larger uncertainties in the individual abundances.
Before discussing an expanded sample of
field stars near the sun, an issue raised in Sec. 3 should be dealt with.
The metallicity distribution  of the young clusters and the young stars
definitively demonstrates that the mean metallicity among recently formed
objects is solar within $\pm$0.1 dex. The cluster sample indicates that
at the $R_{GC}$ of the sun, this mean metallicity is basically the same as
when the sun formed 4.6 Gyr ago. Despite the extraordinary agreement,
analyses of galactic nebular abundances (e.g., Afflerbach {\it et al.} \markcite{a2}1997) 
consistently find that the ISM near the sun is metal-deficient by about
[m/H] = --0.3. It has become commonplace to explain this discrepancy as
evidence that the sun is anomalously metal-rich for its age and location.
The data for the young stars and the young open clusters demonstrate that
if the sun is anomalously metal-rich, so is the typical star formed at $R_{GC}$
= 8.5 kpc over the last 1 Gyr. We suggest
that the source of the discrepancy lies with the nebular abundances, i.e.,
they systematically underestimate the metal content of the HII regions by
about 0.3 dex in the solar neighborhood. Potential problems with nebular
abundances have been under discussion for some time, as evidenced by the
work of Mathis \markcite{m33}(1995) and Alexander \& Balick \markcite{a3}(1997).
However, whether the origin of the proposed deviation
is found within the clouds, within the models, or both is beyond 
the scope of this investigation and the expertise of the investigators.
 
Returning to the field star metallicity distribution, the picture relative
to the young stars and the clusters changes dramatically when one expands the
sample to include any and all field stars near the sun.
The most comprehensive analysis of the local metallicity
distribution to date is that of Wyse \& Gilmore \markcite{w5}(1995; hereafter referred
to as WG). In Fig. 7 we compare the normalized abundance distribution
of the clusters (solid curve) to the Thin Disk (dash-dot curve) and
Thin + Thick Disk (dashed curve) samples of \markcite{w5}WG. The
[Fe/H] scales of the latter two histograms have been offset to
make the curves more distinguishable. It is apparent
that, zero-point uncertainties aside, the field star population in both
groups contains an excess of stars below [Fe/H] = --0.2 which is not
reflected in the cluster sample. The only subset of \markcite{w5}WG which comes close
to reproducing  the cluster distribution is that listed by \markcite{w5}WG as Young Disk.

The disagreement between the field stars and the clusters implies that
they do not sample the same distribution of galactic populations. This
difference is an important clue to the chemical history of the disk
and suggests a straightforward solution. The similarity of the
cluster population to the Young Disk but not the Thin Disk implies 
that the discrepancy arises from differences in the age distributions
of clusters and field stars. The cluster sample contains a
significant fraction of clusters with ages less than 1 Gyr, a rapidly
declining sample with increasing age, and no
clusters older than 9 Gyr; the long-lived field stars provide
a more representative distribution of the entire disk lifetime. 
The discrepancy will arise if field stars with [Fe/H] $<$ --0.2 are 
predominantly members of
the old disk, i.e., they fall in the age range of 8 to 13 Gyr and
do not overlap in age with the surviving cluster population. There was
an open cluster population which overlapped with this field star group
but it has been tidally disrupted.

Such an explanation is consistent with a significant increase in [Fe/H]
within the interstellar medium during the first third of the lifetime
of the galactic disk, followed by a much more gradual increase over the
last 8 Gyr. This trend is in qualitative agreement with the analyses
of field stars in the solar neighborhood as discussed by Twarog \markcite{t7}(1980b)
and revised by Meusinger {\it et al.} \markcite{m21}(1991). (The revision 
by Carlberg {\it et al.}
\markcite{c3}(1985) is invariably cited in discussions of the AMR to illustrate
the changes in the relation as the analysis and the isochrones are
altered and/or improved. Because of biases in the data selection and
analysis, including those discussed by Nissen \markcite{n2}(1995), 
the AMR derived by Carlberg {\it et al.} \markcite{c3}(1985) is unreliable for old disk stars 
and should not be included in the discussion.)

In contrast, though the qualitative 
AMR trends found by Edvardsson {\it et al.} \markcite{e2}(1993; hereafter E93) and
J\o nch-S\o rensen \markcite{j16}(1995) generally agree with the earlier
work, the mean metallicities at a given age are systematically
lower than derived in the earlier work and the ranges in [Fe/H] 
are large enough that stars with [Fe/H]
well below --0.2 can be found at any age beyond 2 Gyr.  In fact, the mean [Fe/H]
at the age of the sun is 
between --0.15 and --0.20 for \markcite{e2}E93, similar to the value
found by J\o nch-S\o rensen \markcite{j16}(1995) for stars near the solar
circle. The mean for the entire sample at the age of the sun for
J\o nch-S\o rensen \markcite{j16}(1995), a sample which is dominated
by stars interior to $R_{GC}$ = 10 kpc, is closer to [Fe/H] = --0.35.
More important, the range in [Fe/H] among field stars 4 Gyr old and older 
for both studies is between 0.6 and 0.8
dex.  Due to the [Fe/H] selection bias, for \markcite{e2}E93 the dispersion is of
questionable value; the range is a more reliable indicator.  
Twarog \markcite{t7}(1980b) 
finds [Fe/H] = --0.05 at 4.6 Gyr
and a dispersion in [Fe/H] near 0.1 dex for for stars younger than the sun.
If the stellar samples in \markcite{e2}E93 and J\o nch-S\o rensen
\markcite{j16}(1995) are representative
of the ISM in the solar neighborhood over the last 10 Gyr, a difference
in the age distribution cannot explain the discrepancy with the clusters.
Clusters with [Fe/H] = --0.2 or lower should be typical of the sample for
ages greater than 2 Gyr and the mean metallicity for clusters greater than
1 Gyr in age should be well below solar. (The mean AMR for the solar
neighborhood as defined by \markcite{e2}E93 is included in the discussion because it
has become a standard reference on the relation despite the [Fe/H] bias
in constructing the sample. The mean abundances with age cannot be
considered reliable (Nissen \markcite{n2}1995)).

If one accepts that the surviving clusters are not atypical of the
ISM in the solar neighborhood, even in the 4 to 8 Gyr range, the resolution
of the discrepancy must reside with the field stars. Before discussing the
probable solution, an often-cited but incorrect option should be eliminated.
In AMR studies F dwarfs are commonly chosen because they evolve on timescales
typical of the lifetime of the galactic disk, a few Gyr. However, while young
stars of any mass that fall within the F-star temperature range are observable,
as a sample ages, the hotter F dwarfs evolve out of the temperature range
and are eliminated. Thus, only the lowest mass F dwarfs will survive over
the entire lifetime of the disk. Because the metallicity and mass of a 
main sequence star determine the star's temperature, the probability of
finding a star of a given [Fe/H] in a temperature-limited sample
changes with age. In general, the limiting age at which one may still
observe a star of a given [Fe/H] declines as [Fe/H] increases. 
Friel \markcite{f11}(1995) cites this selection bias (McClure \& Tinsley
\markcite{m8}1976; Knude 1990)
as the explanation for the existence
of an AMR among the field stars while none occurs within the cluster sample.

However, the selection biases outlined by McClure \& 
Tinsley \markcite{m8}(1976) were well known and were
minimized in the sample selection procedure of Twarog \markcite{t6}(1980a) and
the analysis of the sample by Twarog \markcite{t7}(1980b). All of the effects 
detailed by Knude \markcite{k8}(1990) were modelled and incorporated in the analysis
of the F dwarfs by Twarog \markcite{t5}(1979), as summarized in Twarog
\markcite{t6}\markcite{t7}(1980a,b).
While it is impossible to completely eliminate [Fe/H]-dependent bias
in an F-star sample, the models and analysis in Twarog \markcite{t6}(1980a)
show that for the observed sample, the AMR is not simply a reflection 
of selection
bias. This conclusion is confirmed by the discussion of \markcite{w5}WG who use only
long-lived G dwarfs in their sample, thereby avoiding the possibility of
excluding metal-rich, older stars. The G dwarfs also produce a 
metallicity distribution
heavily weighted toward metallicities lower than the average inner cluster.

\subsection{Galactic Clusters and Field Stars: A Solution}
The fundamental discrepancy is that the metallicity distribution
of the surviving open clusters in the
solar neighborhood at all ages is missing the lower metallicity portion
of the field star distribution. The work of \markcite{e2}E93 and 
J\o nch-S\o rensen \markcite{j16}(1995) implies that the 
metal-weak thin disk can be as young as
2 to 4 Gyr. Where do the lower metallicity field stars come from
if they don't come from the local disk? The answer is supplied by looking
at the cluster sample. While there are no clusters interior to 10 kpc
which overlap with the metal-poor field stars, there is a rich population
of clusters beyond this location which bracket the required [Fe/H]
and age ranges. The large spread in [Fe/H] within the cumulative
cluster sample is often cited as confirmation of the
large range in metallicity derived among the field stars as in \markcite{e2}E93.
But the existence of a discontinuity (or steep gradient) in the disk 
guarantees that the metal-poor clusters beyond $R_{GC}$ have no bearing
on the discussion of the ISM near
the sun and cannot be combined with the local distribution. The property
which makes the clusters different from the field stars isn't the age
but the mass. Though clusters can have galactic orbits which are non-circular,
causing them to move over a range in galactocentric distance as they go
around the galaxy (see, e.g., \markcite{p8}PCA), their collective 
mass ensures that orbital perturbations
caused by the passage of stars and/or massive gas clouds will be negligible.
In contrast individual stars can experience considerable orbital
perturbations from the cumulative effects of such interactions. 
Thus, when combined
with the spatial analysis of Sec. 3, it is concluded that
a significant fraction of the stars in the solar
neighborhood with [Fe/H] $<$ --0.2 are interlopers. These stars are the
field star counterparts to the metal-poor cluster population beyond $R_{GC}$
= 10 kpc. The stars have diffused inward on timescales of a few Gyr, mixing
with the local population to a degree which makes it impossible to distinguish
them dynamically from the locally formed field stars of comparable age.
The remainder of this section will focus on the plausibility of this
solution and how it fits in with the the current picture of the field
star population near the sun.

Before addressing the field star question, a puzzling point regarding
the cluster population should be dealt with. Though one can readily
assume that significant alteration in the galactocentric orbits of the
clusters is unlikely, how much diffusion in galactocentric distance
arises because of the spread in initial conditions among the clusters?
If one takes the orbital analysis by Carraro \& Chiosi \markcite{c26}(1994)
and \markcite{p8}PCA at face value, clusters near the sun typically
range over about 3 kpc in galactocentric distance as they orbit. Under
such conditions, survival of a sharp discontinuity in the galactic
abundance gradient as defined by the clusters seems improbable.

Our clearly biased interpretation of this contradiction is that the
problem lies with the orbits. Neither Carraro \& Chiosi \markcite{c26}(1994)
nor \markcite{p8}PCA give any indication of the uncertainties in their
orbital parameters due to potential errors in distance, radial-velocity,
proper motion, or assumed galactic potential. An indication of the
difficulties in interpreting these kinematic results is available by
comparing the orbits for the five clusters in 
Carraro \& Chiosi \markcite{c26}(1994)
with the same objects in the larger sample of \markcite{p8}PCA. The ranges
in galactocentric distance for clusters in \markcite{p8}PCA were all
significantly larger than found by Carraro \& Chiosi \markcite{c26}(1994);
the {\it increase} in $\Delta$$R_{GC}$ went from a low of 0.9 kpc for NGC 2420 to 4.8 kpc
for NGC 2506. Even more significant is the distribution in orbital
eccentricity, {\it e}, defined as the difference between apogalacticon and
perigalacticon divided by their sum. Of the 19 clusters in \markcite{p8}PCA,
only one has {\it e} below 0.1. If one excludes the extreme case of NGC 7789
({\it e} = 0.6), the remaining 18 clusters have a mean eccentricity of 0.19,
implying an average range near 4 kpc in galactocentric distance for a cluster
at $R_{GC}$ = 10 kpc. No significant difference is found comparing clusters 
sorted by age. While field star samples covering a large range in age 
can exhibit average eccentricities at this level, they are implausible
for an unbiased open cluster sample. The ineffectiveness of cluster
perturbations should lead to average eccentricities only slightly larger
than found among newly formed clusters and stars, i.e., closer to 0.0 than
0.2. Turning the question around, based upon the expected distribution of
{\it e} and the existence of the cluster discontinuity, recent derivations
of the cluster orbits have overestimated the orbital eccentricities
and most clusters diffuse over a much smaller range than
claimed. Even if one chooses a typical value of {\it e} = 0.1 for the clusters,
at $R_{GC}$ = 10 kpc, a cluster will orbit between $R_{GC}$ = 9 kpc and 11 kpc,
enough to round off the edges of the discontinuity but not enough to destroy
it.
 
Since the work of \markcite{e2}E93, a great deal of effort has been expended to
explain the origin of the large dispersion in [Fe/H] among stars of
a given age near the sun; a useful summary of the many options is
given in van den Hoek \& de Jong \markcite{v6}(1997). Out of the many suggestions,
the one of primary relevance is that of Wielen {\it et al.} \markcite{w4}(1995;
hereinafter WFD). 
The goal of WFD was 
to determine the galactocentric
origin of the sun based upon the assumption that over 4.6 Gyr it has
diffused away from its initial location. This can be done if one assumes
that: (a) the original dispersion in [Fe/H] within the ISM at a given age
at a given
galactocentric location was significantly smaller than that observed
among stars more than a few Gyr old; and (b) a significant 
galactocentric gradient in [Fe/H] exists within the disk. If
one knows the AMR, the galactocentric origin of a star of a given age  
can be derived by measuring how much it deviates in [Fe/H] from the
mean value for its age, and moving the star along the galactocentric
abundance gradient by an amount which accounts for the deviation.
Following this approach, \markcite{w4}WFD derive $R_{GC}$ = 6.6 $\pm$0.9 kpc for 
the birthplace of the sun and, more important, conclude that the initial
dispersion in [Fe/H] within the ISM at a given $R_{GC}$ is rather small,
i.e., the range in [Fe/H] found in \markcite{e2}E93 is a product of diffusion.

The solution we propose is qualitatively the same as \markcite{w4}WFD, but has been
modified to account for the results of Sec. 3. Before we discuss those
details, a general comment on the role of stellar diffusion is in order.
The theoretical and observational role of stellar diffusion has been
the focus of numerous investigations over the years (Mayor \markcite{m3}1976; Grenon
\markcite{g5}1987; Francois \& Matteucci \markcite{f8}1993; Fuchs {\it et al.} \markcite{f16}1996, among others).
For the current discussion and that of \markcite{w4}WFD, the most important 
analysis is that by
Wielen \markcite{w3}(1977). Its conclusions are simple but, if correct, devastating to
any investigation which attempts to sort stars into bins of common
origin through kinematics. As summarized in Mihalas \& Binney \markcite{m23}(1981) and
reiterated in \markcite{w4}WFD, the randomizing effects of diffusion on the orbits of
stars, coupled with an imperfect knowledge of the galactic gravitational
potential both in space and time, make the tracing of stellar orbits
of older stars back in time to locate their initial birthplaces a hopeless
task. If the arguments of \markcite{w4}WFD are correct, the conclusion 
of \markcite{e2}E93 that diffusion 
is inadequate to explain the dispersion in [Fe/H] based upon the
approach of Grenon \markcite{g5}(1987) is flawed. A more recent attempt to
constrain the role of diffusion
by van den Hoek \& de Jong \markcite{v6}(1997) is equally questionable on kinematic
grounds and because it depends upon the existence of well-defined radial gradients
in O and Fe.

Assuming that stellar diffusion does occur and the timescales discussed by
Wielen \markcite{w3}(1977) are valid, how does one explain the discrepancy between
the field stars and the clusters? First, over the last 8 Gyr the mean
[Fe/H] of the ISM between 6 and 10 kpc has been solar within $\pm$0.10
dex. Beyond a possible gradual change in [Fe/H] with time, the 
dispersion in [Fe/H] at
all ages has remained $\pm$0.09 dex or smaller. In short,
stars formed in the 6 to 10 kpc range typically have [Fe/H]
between --0.2 and +0.2. Over the same galactocentric
distance range, the abundance gradients in various metals have remained
shallow to flat, i.e., no gradient. Thus, the sun is {\bf not} atypical
of the solar neighborhood 4.6 Gyr ago, a point emphasized in significantly
more detail by \markcite{e2}E93. Because no gradient in [m/H]
exists, the sun could have formed anywhere in the 6 to 10 kpc range.
In general, the approach outlined by \markcite{w4}WFD to identify the birthplace
of field stars becomes moot.

Second, a virtually identical description applies to the galactic disk between
$R_{GC}$ = 10 and 15 kpc, with 
minor modification. The mean [Fe/H] in the outer disk is --0.35, and the
dispersion might be larger, i.e., stars in the range from --0.60 to --0.10
form there. Stars between 6 and 10 kpc diffuse toward $R_{GC}$ = 8.5
kpc, but have no impact on the overall metallicity distribution; they
do change the kinematic distributions with age. However, stars beyond
$R_{GC}$ = 10 kpc also diffuse inward, reaching the solar circle on timescales
of 2 to 3 Gyr (Wielen \markcite{w3}1977). Because the [Fe/H] distribution 
beyond $R_{GC}$ = 10 kpc
is independent of distance, it makes no difference whether a star
comes from $R_{GC}$ = 10.5 kpc or 14.5 kpc.  Because these stars are
kinematically indistinguishable from locally formed stars of similar age, 
they cannot
be isolated from the solar sample and should increase in number
with age. The net result is that the metallicity range among field
stars should remain small among stars younger than 2 to 3 Gyr, but 
increase dramatically beyond this age and remain effectively constant
up to about 8 Gyr. This effect is seen to varying degrees in Twarog
\markcite{t7}(1980b), Meusinger {\it et al.} \markcite{m21}(1991), \markcite{e2}E93, and 
J\o nch-S\o rensen \markcite{j16}(1995). Among the
oldest stars ($>$ 8 Gyr), the exact trend is difficult to predict. If, as
discussed in Sec. 5, the AMR interior to $R_{GC}$ = 10 kpc drops 
between 9 and 12 Gyr ago, the observed dispersion depends upon a variety
of poorly known factors: the ratio of inner disk to outer disk field stars
in the sample, the AMR in the outer disk over the same age range, the
increasing uncertainty in the age determination among older stars, and
the selection bias discussed above which places a cap on the metallicity of
the stars detectable in the oldest age bins. 

\section{A Summary and A Scenario}
An extensive collection of open cluster data has been compiled, standardized,
and merged to permit an analysis of the spatial and chemical properties
of the disk sampled by the cluster population. It is concluded that:

a) The abundance distribution with galactocentric distance for the disk 
clusters is best described by a step function rather than a linear
gradient. Inside $R_{GC}$ = 10 kpc there is, at best, a shallow abundance
gradient which may be the product of the age distribution of the sample.
Beyond 10 kpc, the sample is too small to determine if a comparable
gradient exists.
The discontinuity at 10 kpc separates the cluster sample
into two groups that differ in mean [Fe/H] by about --0.35 dex.

b) There is no evidence for a gradient in abundance perpendicular to the
galactic plane for open clusters. The clusters beyond $R_{GC}$ =
10 kpc included in past discussions have, on average, larger Z distances
that those within 10 kpc. When coupled with a true discontinuity in [Fe/H],
a linear correction with galactocentric distance leaves a residual effect
which translates into a vertical gradient.

c) The metallicity distribution for the clusters is well
described by two Gaussians identified with the
inner and outer clusters. After correcting for a slight abundance 
gradient for the inner clusters, the dispersion in [Fe/H] for the
open clusters at the solar circle reduces to $\pm$0.09 dex while
the average [Fe/H] is approximately solar. Though the
sample of clusters greater than 1 Gyr in age is modest, there is no
indication that either the mean or the dispersion is a significant function
of age. This implies that the metallicity of the sun is not atypical for its age.

d) The metallicity distribution for the inner clusters disagrees with
that derived for field stars in the solar neighborhood. Though the
lack of an abundance gradient removes the specific solution proposed
by \markcite{w4}WFD to explain the large dispersion in [Fe/H] among the field stars,
the general idea of diffusion as the primary culprit survives. The
large metallicity range found among stars with ages greater than 3 Gyr
is the product of diffusion of stars from beyond $R_{GC}$ = 10 kpc.
It should be emphasized that the small dispersion in the cluster metallicity
distribution at the solar circle is independent of the nature of the
galactocentric abundance gradient. The discontinuity, however, provides a
means of increasing the metallicity range among the field stars in the
absence of a linear gradient.

Though the empirically derived trends and the evolutionary scenario are
internally consistent, one key piece of the puzzle is still missing:
why does a discontinuity exist? In reality, this question can be broken
down into two equally difficult problems: how is a discontinuity created,
and how is it maintained over 10 to 12 Gyr?  The solution undoubtedly lies
in understanding why $R_{GC}$ = 10 kpc is so special.
Given the current state of the data and the justifiable uncertainty
regarding the  reality of the discontinuity, a detailed explanation
of the origin of the discontinuity is neither feasible nor warranted.
We can, however, offer a qualitative scenario which fits the observational
evidence at present, both for clusters and field stars.

The primary difference between the inner and outer clusters is a mean
metallicity which differs by 0.3 to 0.4 dex. As noted earlier, there
is no evidence that the mean abundance in either zone has changed
significantly over the last 8 Gyr though, again, increases on the order
of 0.1 dex cannot be excluded. In most models of galactic chemical
evolution, as time passes the mean metallicity of the gas out of which
the stars form will approach some limiting value in [m/H], in large part
because of the logarithmic nature of [m/H]. The limiting value is some
fraction of the nucleosynthetic yield; the exact value of the
fraction depends upon the stellar yields, the initial mass 
function, gas flows, and the
element under consideration. Thus, it is possible that the effective
yield for the chemical evolution of the outer disk is simply lower
than that of the inner disk for some currently unknown reason.
As an alternative, we offer the following scenario:

The discontinuity at $R_{GC}$ = 10 kpc is a reflection of the
original boundary of the newly formed disk, currently referred
to as the thick disk. Chemically and kinematically, the thick
disk is associated with the disk globular cluster population
typified by 47 Tuc and M71, and extends to, at least, the solar
circle (Armandroff \markcite{a12}1993; Zinn \markcite{z1}1996 and references therein). Through
the use of the rich population of open clusters, the exact
radial boundary of the thick disk is set at $R_{GC}$ = 10 kpc. As outlined
by \markcite{w5}WG based primarily upon the data of \markcite{e2}E93, this population originated 
about 12 Gyr ago when the typical
metallicity of the gas near $R_{GC}$ = 10 kpc was [Fe/H] $\sim$ --1.0.
It is assumed that the thick disk evolves dynamically and chemically 
as a separate entity from the surrounding halo in that while gas may infall
from the halo to the disk, the evolution of the thick disk has little
impact on the surrounding halo. While the metallicity within 
the thick disk rises from
[Fe/H] $\sim$ --1.0 to --0.6, the mean metallicity in the surrounding halo
remains relatively unchanged. Between 10 and 8 Gyr ago, the mean [m/H]
of the inner disk increased rapidly from --0.6 to approximately --0.1 and
the inner disk evolved from the thick disk to the standard thin disk. How
quickly this occured remains unknown, but by 8 Gyr ago, the mean
[m/H] was close to solar. Over the same time scale the outer disk 
takes shape, and chemical evolution follows a similar trend. However, the
initial metallicity of the outer disk remained [m/H] $\sim$ --1.0
because of the lack of the thick disk phase that drove the chemical
enrichment of the inner disk.  In the long run this produced a difference in
the current [m/H] of the inner and outer disks which reflects the offset
between the typical thick disk star ([m/H] = --0.6 to --0.7) and the halo
transition ([m/H] = --1.0).

If nothing additional occurred, one should expect the metallicity distribution
near the sun to reflect three populations: thick disk with [m/H] between
--1.0 and --0.5, thin disk with [m/H] between --0.3 and +0.2, and a transition
population connecting the two. The size of the transition population depends
upon the timescale over which it evolved and the star formation rate during
this period, neither of which are known. This rather classic picture of
disk formation is distorted by the additional factor of diffusion from
the outer disk. Because the mean [m/H] in the outer zone has remained 
constant for such a long time, the metallicity distribution of the diffused
stars is dominated by stars in the [m/H] range from --0.1 to --0.6, 
exactly the range occupied by the transition population, and merges
locally to become what \markcite{w5}WG refer to as the metal-weak thin disk.

For the additional question of the survival of the discontinuity, we have
no explanation. If the initial metallicities of the gas in the outer and
inner disk were systematically different, one would expect radial gas diffusion
to play a role in smearing, if not wiping out, the discontinuity over the
timescale of a few Gyr. Yet the discontinuity has survived until at least
1 Gyr ago. The potential discontinuity at $R_{GC}$ = 6 kpc proposed by
Simpson {\it et al.} \markcite{s13}(1995) is located at the outer edge of a ring of
molecular clouds and associated spiral arms; they suggest that the
step function is due to radial mixing driven by the presence of a
bar. This suggestion has no direct bearing on what happens at 10 kpc
except to highlight the need for a dynamical boundary separating the
gas in the inner and outer zones. Potential evidence for some dynamical
difference between the inner and outer zones is supplied by the scale
height of the clusters. 

Though the sample is by no means complete and the plot includes
clusters younger than 1 Gyr in the inner zone, the Z distribution of clusters
in Fig. 5 suggests that the scale height of the old cluster population in
the outer zone may differ from that in the inner zone. Janes \& Phelps
\markcite{j9}(1994) have analyzed the Z distribution of the old clusters and derived
a scale height of 325 pc, significantly larger than the 55 pc estimate from young
open clusters. Because the ages of the older sample range from 1 to 9
Gyr and diffusion in the Z direction will not work for clusters, one is
left with tidal disruption of clusters near the galactic plane or 
satellite mergers to explain the apparently large scale height. The former
explanation is rejected because any attempt to explain the large-Z clusters
as the high-Z tail of the thin disk population overproduces the observable
number of clusters near the plane, even accounting for tidal disruption.
Thus, the old cluster Z-distribution implies evidence for regular mergers
over the last 10 Gyr.

Given the discontinuity in the disk, an alternate interpretation comes
to mind. A fundamental assumption of the Janes \& Phelps \markcite{j9}(1994) analysis
is that all the old clusters come from the same population. Thus, when a
scale height is derived, it is applicable to the cluster count at all
galactocentric radii. If the discontinuity at 10 kpc applies to scale
height as well, i.e., the larger Z range for clusters beyond 10 kpc in
Fig. 5 is real, the scale height of 325 pc derived by Janes
\& Phelps \markcite{j9}(1994) is a composite. For the cluster sample interior to $R_{GC}$
= 10 kpc, the true observed scale height is intermediate to 55 pc and 325 pc,
the product of an original scale height near 55 pc, altered
by tidal disruption over time through preferential destruction of
the clusters near the plane. Beyond $R_{GC}$ = 10 kpc, the scale 
height is unknown; we have no real information on the
scale height of old clusters beyond 10 kpc because there is little
reliable information on the number of clusters near the galactic plane.
While those away from the plane are easily discovered, those within the
disk remain hidden by dust or within a rich background field of stars; 
this is apparent if one compares the radial distribution
of old and young clusters (Figs. 7 and 9) in Janes \& Phelps \markcite{j9}(1994) for
$R_{GC}$ $>$ 10 kpc. If the scale height of the young disk in the outer zone
or, more important, of the disk at the time of formation of the older clusters
is greater than 55 pc, when combined with selection effects
and the lower probability of 
tidal disruption at larger galactocentric distance, the excess of high-Z
clusters can be reduced.
This still implies that the true scale height of the outer disk is greater than
that of the inner disk; additionally, it is likely that cluster
disruption is more efficient inside $R_{GC}$ = 10 kpc. If either suggestion
can be confirmed observationally, it would support the notion of a
dynamical difference between the inner and outer galaxy.  Note that the
suggestion of a thicker disk in the outer galaxy is also consistent with
the explanation of mergers as a thickening agent in galactic evolution
(Janes \& Phelps \markcite{j9}1994).

We close this paper on a somewhat pessimistic, but debatable, note.
A primary goal of field star studies is to delineate the chemical evolution
of the disk at a particular location over time. All the studies of the
AMR to date have been premised on the assumption that one can isolate a field
star sample ranging in age over the lifetime of the disk which typifies the
ISM at the solar circle over that same timescale, i.e., it has been
assumed that stellar diffusion did not mix stars from significantly
different galactocentric origins. If the arguments of Wielen \markcite{w3}(1977) and
\markcite{w4}WFD are correct, this assumption fails and kinematics cannot be used to
sort stars indivdually into bins of galactocentric origin. With the existence
of a linear abundance gradient, \markcite{w4}WFD showed that the metallicity could be
used to distinguish among stars formed at various galactocentric
distances. If the cluster analysis above is correct, this positional
tag is eliminated. The net result is that unless an alternative method
is derived for isolating the metal-weak thin disk stars which come from
beyond $R_{GC}$ = 10 kpc from those which formed locally as part of the
transition between the thick and the thin disk, any attempt to understand
the chemical and dynamical history of the disk at the solar circle using
field stars may remain an exercise in futility. One possibility is that at a
given [Fe/H],
the outer zone stars might exhibit abundance ratios, e.g., [O/Fe] or [Ca/Fe],
which separate them from
the inner sample. Clusters represent the ideal object but any clusters
formed in the thick disk have long since been destroyed and any distant
clusters in this age range well beyond the solar circle tell us nothing
about the local galactic disk.

It is a pleasure to acknowledge the help of B. Carney,
E. Friel, A. Fry,
M. Gim, E. Hufnagel, R. D. McClure, and J. Shields
who supplied information and/or comments which aided this investigation.
The clarity of the paper has been improved thanks to the thoughtful
comments of the referee. This research has made use of the SIMBAD
database, operated at CDS, Strasbourg, France.

\newpage
\figcaption{The absolute magnitude of the red giant clump as a function of
(a) metallicity and (b) age. For Fig. 1b, cluster identifications are aligned
vertically with the points.}
\figcaption{Metallicity differences between the revised DDO abundances and
those of PCA (open circles) and FJ (crosses), in the 
sense (REF - DDO). The
data of FJ have been shifted by +0.05 in [Fe/H] to place 
them on the PCA scale.}
\figcaption{(a) Cluster abundances as a function of galactocentric position
as given in Table 2. (b) Same as (a) without error bars. Open circles 
are rederived DDO abundances, open triangles are spectroscopic abundances,
closed triangles are combined DDO and spectroscopic results, and open
squares are transformed DDO abundances from unpublished photometry.}
\figcaption{Absolute distance away from the galactic plane as a function
of metallicity for the clusters used by PCA (open circles) and the
additional sample of FJ (crosses).}
\figcaption{Absolute distance away from the plane as a function of
galactocentric distance. Same symbols as Fig. 4.}
\figcaption{Same as Fig. 4 after adjusting the clusters beyond $R_{GC}$ =
10 kpc for an offset in [Fe/H] of 0.32.}
\figcaption{Metallicity distribution for field stars, Thin + Thick Disk
(dashed curve), Thin Disk (dash-dot curve) from WG, and the inner clusters
(solid curve). The first two histograms have been offset by 0.02 in [Fe/H]
from the solid curve to make the curves more distinguishable.}

\begin{references}
\reference{a1} Adler, D. S., \& Janes, K. A. 1982, \pasp, 94, 905
\reference{a2} Afflerbach, A., Churchwell, E., \& Werner, M. W. 1997, \apj, 478, 190
\reference{a3} Alexander, J., \& Balick, B. 1997, \aj, 114, 713
\reference{a4} Anthony-Twarog, B. J., Kaluzny, J., Shara, M. M., \& 
Twarog, B. A. 1990, \aj, 99, 1504
\reference{a5} Anthony-Twarog, B. J., Mukherjee, K., Caldwell, N., \& 
Twarog, B. A. 1988, \aj, 95, 1453
\reference{a6} Anthony-Twarog, B. J., Payne, D. M., \& Twarog, B. A. 1989a, 
\aj, 97, 1048
\reference{a7} Anthony-Twarog, B. J., \& Twarog, B. A. 1985, \apj, 291, 595
\reference{a8} Anthony-Twarog, B. J., \& Twarog, B. A. 1987, \aj, 94, 1222
\reference{a9} Anthony-Twarog, B. J., Twarog, B. A., \& McClure, R. D. 1979, 
\apj, 233, 188
\reference{a10} Anthony-Twarog, B. J., Twarog, B. A., \& Sheeran, M. 1994, 
\pasp, 106, 486
\reference{a11} Anthony-Twarog, B. J., Twarog, B. A., \& Shodhan, S. 1989b, 
\aj, 98, 1634
\reference{a12} Armandroff, T. E. 1993, in The Globular Clusters
- Galaxy Connection, ASP Conf. Ser. 48,  edited by  G. H. Smith and 
J. P. Brodie (ASP, San Francisco), p. 48
\reference{a13} Arp, H. C., \& Cuffey, J. 1962, \aj, 136, 51
\reference{a14} Ashman, K. M., Bird, C. M., \& Zepf, S. E. 1994, \aj, 108, 2348
\reference{a15} Auner, G. 1974, \aaps, 13, 143
\reference{b1} Becker, W. \& Fenkart, R. 1971, \aaps, 4, 241
\reference{b2} Beers, T. C., Flynn, K., \& Gebhardt, K. 1990, \aj, 100, 32
\reference{b3} Bergbusch, P. A., VandenBerg, D. A., \& Infante, L. 1991, \aj, 101, 2102
\reference{b4} Bird, C. M., \& Beers, T. C. 1993, \aj, 105, 1596
\reference{b5} Boesgaard, A. M. 1989, \apj, 336, 798
\reference{b6} Bonifazi, A., Fusi-Pecci, F., Romeo, G., \& Tosi, M. 1990, 
\mnras, 245, 15
\reference{b7} Brocato, E., Castellani, V., \& DiGiorgio, A. 1993, \aj, 105, 2192 
\reference{b8} Brosterhus, E. 1963, Astron. Ahb. Hamb. Sternw. VII No. 2
\reference{b9} Brown, J. A., Wallerstein, G., Geisler, D., \& Oke, J. B. 1996, 
\aj, 112, 4
\reference{b10} Burkhead, M. S. 1969, \aj, 74, 1171
\reference{b11} Burkhead, M. S. 1971, \aj, 76, 251
\reference{b12} Burkhead, M. S., Burgess, R. D., \& Haisch, B. M. 1972, 
\aj, 77, 661
\reference{c1} Cameron, L. M. 1985, \aap, 147, 47
\reference{c2} Cannon, R. D., \& Lloyd, C. 1969, \mnras, 144, 449
\reference{c3} Carlberg, R., Dawson, P., Hsu, T., \& VandenBerg, D. 1985, \apj, 294,
674
\reference{c4} Carraro, G., \& Chiosi, C. 1994, \aap, 287, 761 
\reference{c26} Carraro, G., \& Chiosi, C. 1995, \aap, 288, 751
\reference{c5} Carraro, G., Chiosi, C., Bressan, A., \& 
Bertelli, G. 1994, \aaps, 103, 375
\reference{c6} Carraro, G., \& Patat, F. 1995, \mnras, 276, 563
\reference{c7} Chincarini, G. 1963, Contr. Asiago No. 138
\reference{c8} Christian, C. A. 1981, \apj, 246, 827
\reference{c9} Christian, C. A. 1984, \apj, 286, 552
\reference{c10} Christian, C. A., Heasley, J. N., \& Janes, K. A. 1985, \apj, 299, 683
\reference{c12} Clari\'{a}, J. J. 1973, \aaps, 9, 251
\reference{c13} Clari\'{a}, J. J. 1980, \apss, 72, 347
\reference{c14} Clari\'{a}, J. J. 1982, \aaps, 47, 323
\reference{c15} Clari\'{a}, J. J. 1985, \aaps, 59, 195
\reference{c16} Clari\'{a}, J. J., \& Lapasset, E. 1983, JAp\&A, 4, 117
\reference{c17} Clari\'{a}, J. J., \& Lapasset, E. 1986, \apj, 302, 656
\reference{c18} Clari\'{a}, J. J., \& Lapasset, E. 1988, \mnras, 235, 1129
\reference{c19} Clari\'{a}, J. J., \& Lapasset, E. 1989, \mnras, 241, 301
\reference{c20} Clari\'{a}, J. J., Lapasset, E., \& Minniti, D. 1989, \aaps, 78, 363
\reference{c21} Clari\'{a}, J. J., \& Mermilliod, J. -C. 1992, \aaps, 95, 429
\reference{c22} Clari\'{a}, J. J., Mermilliod, J. -C., Piatti, A. E., 
\& Minniti, D. 1994, \aaps, 107, 39
\reference{c23} Clari\'{a}, J. J., Piatti, A. E., \& Osborn, W. 1996, \pasp, 108, 672
\reference{c24} Crinklaw, G., \& Talbert, F. D. 1991, \pasp, 103, 536
\reference{c25} Cudworth, K. \& Anthony-Twarog, B. J. 1997, private communication
\reference{d8} Dachs, J., \& Kabus, H. 1989, \aaps, 78, 25
\reference{d1} Daniel, S. A., Latham, D. W., Mathieu, R. D., \& Twarog, B. A. 1994,
\pasp, 106, 281
\reference{d2} Dawson, D. 1978, \aj, 83, 1424
\reference{d3} Dawson, D. 1981, \aj, 86, 237
\reference{d4} Deming, D., Olson, E. C., \& Yoss, K. M. 1977, \aap, 57, 417
\reference{d5} Dinerstein, H. L., Haas, M. R., Erickson, E. F., \& Werner, M. W.
1993, \baas, 25, 850
\reference{d6} Dinescu, D. I., Girard, T. M., van Altena, W. F., Yang, T.-G., \&
Lee, Y.-W. 1996, \aj, 111, 1205
\reference{d7} Dodd, R. J., MacGillivray, H. T., \& Hilditch, R. W. 1977, 
\mnras, 181, 729
\reference{e1} Ebbighausen, E. G. 1939, \apj, 90, 689
\reference{e2} Edvardsson, B., Andersen, J., Gustaffson, B., Lambert, D. L., 
Nissen, P. E., \& Tomkin, J. 1993, \aap, 275, 101 (E93)
\reference{e3} Eggen, O. J. 1968, \apj, 152, 83
\reference{e4} Eggen, O. J. 1969, \apj, 155, 439
\reference{e5} Eggen, O. J. 1972, \apj, 173, 63
\reference{e6} Eggen, O. J. 1980, \apj, 238, 627
\reference{e7} Eggen, O. J. 1983, \aj, 88, 184
\reference{e8} Eggen, O. J. 1989, \pasp, 101, 54
\reference{f1} Feinstein, A., Cabrera, A. L., \& Clari\'{a}, J. J. 1978, \aaps, 43, 241
\reference{f2} Feinstein, A., \& Forte, J. C. 1974, \pasp, 86, 284
\reference{f3} Fenkart, M. P., Buser, R., Ritter, H., Schmitt, H., Steppe, H., 
Wagner, R., \& Wiedmann, D. 1972, \aaps, 7, 487
\reference{f4} Fernandez, J. A., \& Salgado, C. W. 1980, \aaps, 39, 11
\reference{f5} Fernie, J. D. 1963, \aj, 68, 780
\reference{f6} Fich, M., \& Silkey, M. 1991, \apj, 366, 107
\reference{f7} Francic, S. P. 1989, \aj, 98, 888
\reference{f8} Francois, P., \& Matteucci, F. 1993, \aap, 280, 136
\reference{f9} Frandsen, S., Balona, L. A., Viskum, M., Koen, C., \& Kjeldsen, H. 1996,
\aap, 308, 132
\reference{f10} Friel, E. D. 1989, \pasp, 101, 244
\reference{f11} Friel, E. D. 1995, \araa, 33, 381 
\reference{f12} Friel, E. D. 1997, private communication
\reference{f13} Friel, E. D., \& Janes, K. A. 1993, \aap, 267, 75 (FJ)
\reference{f14} Frogel, J. A., \& Twarog, B. A. 1983, \apj, 274, 270
\reference{f15} Fry, A. M., \& Carney, B. W. 1997, \aj, 113, 1073
\reference{f16} Fuchs, B., Dettbarn, C., \& Wielen, R. 1996, in
Unsolved Problems of the Milky Way, IAU Symp. No. 169, edited by L. Blitz 
and P. Teuben (Kluwer Academic, Dordrecht), p. 431  
\reference{g1} Geisler, D. P., \& Smith, V. V. 1984, \pasp, 96, 871
\reference{g2} Gim, M. 1997, private communication
\reference{g3} Girard, T. M., Grundy, W. M., Lopez, C. E., \& van Altena, 
W. F. 1989, \aj, 98, 227
\reference{g4} Glushkova, E. V., \& Rastorguev, A. S. 1991, Soviet Astron. 
Lett., 17, 13
\reference{g5} Grenon, M. 1987, JAp\&A, 8, 123
\reference{h1} Hardy, E. 1979, \aj, 84, 319
\reference{h2} Harris, G. L. H. 1976, \apjs, 30, 451
\reference{h3} Harris, G. L. H., Fitzgerald, M. P. V., 
Mehta, S., \& Reed, B. C. 1993, \aj, 106, 1533
\reference{h4} Harris, G. L. H., \& Harris, W. E. 1977, \aj, 82, 612
\reference{h5} Harris, H. C., \& McClure, R. D. 1985, \pasp, 97, 261
\reference{h6} Hartwick, F. D. A., \& Hesser, J. E. 1973, \apj, 183, 883
\reference{h7} Hartwick, F. D. A., Hesser, J. E., \& McClure, R. D. 1972, \apj, 174, 557
\reference{h8} Hartwick, F. D. A., \& McClure, R. D. 1972, \pasp, 84, 288
\reference{h9} Hassan, S. M. 1976, \aaps, 26, 13
\reference{h10} Hawarden, T. G. 1975, \mnras, 173, 801
\reference{h11} Hawarden, T. G. 1976a, \mnras, 174, 225
\reference{h12} Hawarden, T. G. 1976b, \mnras, 174, 471
\reference{h13} Herzog, A. D., Sanders, W. L., \& Seggewiss, W. 1975, \aaps,
19, 211
\reference{h14} Hesser, J. E., \& Smith, G. H. 1987, \pasp, 99, 1044
\reference{h15} Hiltner, W. A., Iriarte, B., \& Johnson, H. L. 1958, \apj, 127, 539
\reference{h16} Hirshfeld, A., McClure, R. D., \& Twarog, B. A. 1978, in
The HR Diagram, IAU Symp. No. 80, edited by A. G. D. Philip and D. S. 
Hayes (Reidel, Dordrecht), p. 163
\reference{h17} Hoag, A. A., Johnson, H. L., Iriarte, B., Mitchell, R. J., Hallam, K. L.,
\& Sharpless, S. 1961, Publ. U.S. Naval Obs., 17, 349
\reference{h18} Hobbs, L. M., Thorburn, J. A., \& Rodriguez-Bell, T. 1990, \aj, 100, 710
\reference{h19} Houdeshelt, M. L., Frogel, J. A., \& Cohen, J. G. 1992, \aj, 103, 163
\reference{i1} Ianna, P. A., Adler, D. S., \& Faudree, E. F. 1987, \aj, 93, 347
\reference{j1} Jahn, K., Kaluzny, J., \& Rucinski, S. M. 1995, \aap, 295, 101
\reference{j2} Janes, K. A. 1975, \apjs, 29, 161
\reference{j3} Janes, K. A. 1977a, \aj, 82, 35
\reference{j4} Janes, K. A. 1977b, \pasp, 89, 576
\reference{j5} Janes, K. A. 1979, \apjs, 39, 135
\reference{j6} Janes, K. A. 1981, \aj, 86, 1210
\reference{j7} Janes, K. A. 1984, \pasp, 96, 977
\reference{j8} Janes, K. A., \& Adler, D. 1982, \apjs, 49, 425
\reference{j9} Janes, K. A., \& Phelps, R. L. 1994, \aj, 108, 1773
\reference{j10} Janes, K. A., \& Smith, G. H. 1984, \aj, 89, 487
\reference{j11} Jankowitz, N. E., \& McCosh, C. J. 1963, MNASSA, 22, 18
\reference{j12} Jennens, P. A., \& Helfer, H. L. 1975, \mnras, 172, 701
\reference{j13} Johnson, H. L. 1952, \apj, 116, 640 
\reference{j14} Johnson, H. L., Hoag, A. A., Iriarte, B., Mitchell, R. I., \& 
Hallam, K. L. 1961, Lowell Obs. Bull., No. 113
\reference{j15} Johnson, H. L., Sandage, A. R., \& Wahlquist, H. D. 1956, \apj, 124, 81
\reference{j16} J\o nch-S\o rensen, H. 1995, \aap, 298, 799
\reference{k1} Kaluzny, J. 1988, Acta Astr., 38, 339
\reference{k2} Kaluzny, J., Krzeminski, W., \& Mazur, B. 1996, \aaps, 118, 303
\reference{k3} Kaluzny, J., Mazur, B., \& Krzeminski. W. 1993, \mnras, 262, 49
\reference{k4} Kaluzny, J., \& Richtler, T. 1989, Acta Astr., 29, 139
\reference{k5} Kaluzny, J. \& Rucinski, S. M. 1995, \aaps, 114, 1
\reference{k6} Kassis, M., Janes, K. A., Friel, E. D., \& 
Phelps, R. L. 1997, \aj, 113, 1723
\reference{k7} Kjeldsen, H., \& Frandsen, S. 1991, \aaps, 87, 119
\reference{k8} Knude, J. 1990, \aap, 230, 16
\reference{k9} Kozhurina-Platais, V., Demarque, P., Platais, I., Orosz, J. A., \&
Barnes, S. 1997, \aj, 113, 1045
\reference{k10} Kozhurina-Platais, V., Girard, T. M., Platais, I., van Altena, W. F.,
Ianna, P., \& Cannon, R. D. 1995, \aj, 109, 672
\reference{k11} Kubiak, M., Kaluzny, J., Krzeminski, W., \& Mateo, M. 1992, 
Acta Astr., 42, 155
\reference{l1} Larsson-Leander, G. 1964, \apj, 140, 144
\reference{l2} Lee, C. W., Mathieu, R. D., \& Latham, D. W. 1989, \aj, 97, 1710
\reference{l3} Lindholm, E. N. 1957, \apj, 126, 588
\reference{l4} Lindoff, U. 1968, Ark. Astron., 5, 63
\reference{l5} Lindoff, U. 1972a, \aaps, 7, 231
\reference{l6} Lindoff, U. 1972b, \aaps, 7, 497
\reference{l7} Lindoff, U., \& Johansson, K. 1968, Ark. Astron., 5, 45
\reference{l8} Lohmann, W. 1961, AN, 286, 105
\reference{l9} Luck, R. E. 1982, \apj, 256, 177
\reference{l10} Luck, R. E. 1991, \apjs, 75, 579
\reference{l11} Lyng\aa, G. 1987, Catalogue of Open Clusters, Center de Donnees 
Stellaires, Strasbourg
\reference{m1} Maciel, W. J., \& Koppen, J. 1994, \aap, 282, 436
\reference{m2} Mathieu, R. D., Latham, D. W., Griffin, R. F., \& Gunn, J. E. 
1986, \aj, 92, 1100
\reference{m33} Mathis, J. S. 1995, RMxA\&A, Conference Series, 3, 207
\reference{m3} Mayor, M. 1976, \aap, 48, 301
\reference{m4} McClure, R. D. 1972, \apj, 172, 615
\reference{m5} McClure, R. D. 1974, \apj, 194, 355
\reference{m6} McClure, R. D. 1997, private communication
\reference{m32} McClure, R. D., \& Forrester, W. 1981, Pub. DAO, No. 14, 439
\reference{m7} McClure, R. D., Forrester, W. T., \& Gibson, J. 1974, \apj, 189, 409
\reference{m8} McClure, R. D., \& Tinsley, B. M. 1976, \apj, 208, 480
\reference{m9} McClure, R. D., Twarog, B. A., \& Forrester, W. T. 1981, \apj, 243, 841
\reference{m10} McLachlan, G. J., \& Basford, K. E. 1988, in Mixture Models:
Inference and Applications to Clustering, (Marcel Dekker, New York)
\reference{m11} McNamara, B. J., \& Solomon, S. J. 1981, \aaps, 43, 337
\reference{m12} McWilliam, A. 1990, \apjs, 74, 1075 
\reference{m14} Mermilliod, J. -C. 1981, \aaps, 44, 467
\reference{m15} Mermilliod, J. -C., Andersen, J., Nordstrom, B., \& Mayor, M. 
1995, \aap, 299, 53
\reference{m16} Mermilliod, J. -C., Huestamendia, G., \& del Rio, G. 1994,
\aaps, 106, 419
\reference{m17} Mermilliod, J. -C., Huestamendia, G., del Rio, G., \& 
Mayor, M. 1996, \aap, 307, 80
\reference{m18} Mermilliod, J. -C., \& Mayor, M. 1989, \aap, 219, 125
\reference{m19} Mermilliod, J. -C., \& Mayor, M. 1990, \aap, 237, 61
\reference{m20} Mermilliod, J. -C., Mayor, M., \& Burki, G. 1987, \aaps, 70, 389
\reference{m21} Meusinger, H., Reimann, H. -G., \& Stecklum, B. 1991, \aap, 245, 57
\reference{m22} Meynet, G., Mermilliod, J. -C., \& Maeder, A. 1993, \aaps, 98, 477
\reference{m23} Mihalas, D., \& Binney, J. 1981, in Galactic Astronomy, (W.H. Freeman,
San Francisco), p. 437
\reference{m24} Minniti, D. 1995, \aaps, 113, 299
\reference{m25} Moffat, A. F. J., \& Vogt, N. 1973, \aaps, 10, 135
\reference{m26} Moffat, A. F. J., \& Vogt, N. 1975, \aaps, 20, 85
\reference{m27} Moll\'{a}, M., Ferrini, F., \& Diaz, A. I. 1997, \apj, 475, 519
\reference{m28} Montgomery, K. A., Marschall, L. A., \& Janes, K. A. 1993, \aj, 106, 181
\reference{m29} Montgomery, K. A., Janes, K. A., \& Phelps, R. L. 1994, \aj, 108, 585
\reference{m30} Morrison, H. L., Flynn, C. M., \& Freeman, K. C. 1990, \aj, 100, 1191
\reference{m31} Murray, R. L., Anthony-Twarog, B. J., \& Twarog, B. A. 1988,
\baas, 20, 717
\reference{n1} Nissen, P. E. 1988, \aap, 199, 146
\reference{n2} Nissen, P. E. 1995, in Stellar Populations, IAU Symp. No. 164,
edited by P. C. van der Kruit and G. Gilmore (Kluwer Academic, Dordrecht), p. 109
\reference{n3} Nordstr\"{o}m, B., Andersen, J., \& Andersen, M. I. 1997, \aap, 322, 460
\reference{n4} Noriega-Mendoza, H., \& Ruelas-Mayorga, A. 1997, \aj, 113, 722
\reference{n5} Norris, J., \& Hawarden, T. G. 1978, \apj, 223, 483
\reference{p1} Panagia, N., \& Tosi, M. 1981, \aap, 96, 306
\reference{p2} Pastoriza, M. G., \& R\"{o}pke, U. O. 1983, \aj, 88, 1769
\reference{p3} Pedreros, M. 1987, \aj, 94, 1237
\reference{p4} Pe\~{n}a, J. H., \& Peniche, R. 1994, RMxA\&A, 28, 139
\reference{p5} Pe\~{n}a, J. H., Peniche, R., Bravo, H., \& Yam, O. 1994, 
RMxA\&A, 28, 7
\reference{p6} Pesch, P. 1961, \apj, 134, 602
\reference{p7} Phelps, R. L., Janes, K. A., \& Montgomery, K. A. 1994, \aj, 107, 1079
\reference{p8} Piatti, A. E., Clari\'{a}, J. J., \& Abadi, M. G. 1995, \aj, 110, 2813 (PCA)
\reference{p9} Piatti, A. E., Clari\'{a}, J. J., \& Minniti, D. 1993, JAp\&A,
14, 145
\reference{p10} Platais, I. 1991, \aaps, 87, 577
\reference{p11} Prosser, C. F., Stauffer, J. R., Caillault, J. -P., Balachandran, S.,
Stern, R. A., \& Randich, S. 1995, \aj, 110, 1229
\reference{r1} Ramsay, G., \& Pollaco, D. L. 1992, \aaps, 94, 73
\reference{r2} Ratnatunga, K. U., \& Freeman, K. C. 1989, \apj, 339, 126
\reference{r3} Richtler, T., \& Kaluzny, J. 1989, \aaps, 81, 225
\reference{r4} Rosvick, J. M. 1995, \mnras, 277, 1379
\reference{s1} Sagar, R., \& Sharples, R. M. 1991, \aaps, 88, 47
\reference{s2} Sandage, A. R., \& Fouts, G. 1987, \aj, 93, 74
\reference{s3} Sanders, W. L. 1990, \aaps, 84, 615
\reference{s4} Sanders, W. L., \& Schr\"{o}der, R. 1980, \aap, 88, 102
\reference{s5} Schmidt, E. G. 1976, \pasp, 88, 63
\reference{s6} Schmidt, E. G. 1978, \pasp, 90, 157
\reference{s7} Schmidt, E. G. 1982, \pasp, 94, 232
\reference{s8} Schmidt, E. G. 1984, \apjs, 55, 455
\reference{s9} Schmidt-Kaler, T. 1961, AN, 286, 113
\reference{s10} Scott, J. E., Friel, E. D., \& Janes, K. A. 1995, \aj, 109, 1706
\reference{s11} Shaver, P. A., McGee, R. X., Newton, L. M., Danks, A. C., \&
Pottasch, S. R. 1983, \mnras, 204, 53
\reference{s12} Shobbrook, R. R. 1986, \mnras, 220, 825
\reference{s13} Simpson, J. P., Colgan, S. W. J., Rubin, R. H., Erickson, E. W., 
\& Haas, M. R. 1995, \apj, 444, 721
\reference{s14} Smartt, S. J., \& Rolleston, W. R. J. 1997, \apjl, 481, L47
\reference{s15} Smith, G. E. 1982, \aj, 87, 360
\reference{s16} Smith, G. E. 1983, \pasp, 95, 296
\reference{s17} Smyth, M. J., \& Nandy, K. 1962, Publ. R. Obs. Edinburgh, 3, 21
\reference{s18} Stetson, P. B. 1981, \aj, 86, 1500
\reference{t1} Taylor, B. J. 1991, \apjs, 76, 715
\reference{t2} Thogersen, E. N., Friel, E. D., \& Fallon, V. 1993, \pasp, 105, 1253
\reference{t3} Torres, G., Stefanik, R. P., \& Latham, D. W. 1997, \apj, 474, 256
\reference{t4} Twarog, B. A. 1978, \apj, 220, 890
\reference{t5} Twarog, B. A. 1979, Ph.D. Thesis, Yale University
\reference{t6} Twarog, B. A. 1980a, \apjs, 44, 1
\reference{t7} Twarog, B. A. 1980b, \apj, 242, 242
\reference{t8} Twarog, B. A. 1981, \aj, 86, 386
\reference{t9} Twarog, B. A. 1983, \apj, 267, 207
\reference{t10} Twarog, B. A., \& Anthony-Twarog, B. J. 1989, \aj, 97, 759
\reference{t11} Twarog, B. A., \& Anthony-Twarog, B. J. 1996, \aj, 112, 1500
\reference{t12} Twarog, B. A., Anthony-Twarog, B. J., \& Hawarden, T. G. 1995, 
\pasp, 107, 1215
\reference{t13} Twarog, B. A., Anthony-Twarog, B. J., \& McClure, R. D. 1993, 
\pasp, 105, 78
\reference{t14} Twarog, B. A., \& Tyson, N. 1985, \aj, 90, 1247
\reference{v1} VandenBerg, D. A. 1985, \apjs, 58, 711
\reference{v2} VandenBerg, D. A., \& Poll, H. E. 1989, \aj, 98, 1451
\reference{v3} van den Bergh, S. 1977, \apj, 215, 89
\reference{v4} van den Bergh, S., \& Heeringa, R. 1970, \aap, 9, 209
\reference{v5} van den Bergh, S., \& McClure, R. D. 1980, \aap, 88, 360
\reference{v6} van den Hoek, L. B., \& de Jong, T. 1997, \aap, 318, 231
\reference{v7} Vansevicius, V., Platais, I., Paupers, O., \& Abolins, E. 1997, 
\mnras, 285, 871
\reference{v10} Venn, K. A. 1995, \apjs, 99, 659
\reference{v8} Vidal, N. V. 1973, \aaps, 11, 93
\reference{v9} Vogt, N., \& Moffat, A. F. J. 1973, \aaps, 9, 97
\reference{w1} Walker, A. R. 1985, \mnras, 214, 45
\reference{w2} West, F. R. 1967, \apjs, 14, 359
\reference{w3} Wielen, R. 1977, \aap, 60, 263
\reference{w4} Wielen, R., Fuchs, B., \& Dettbarn, C. 1996, \aap, 314, 438 (WFD)
\reference{w5} Wyse, R. F. G., \& Gilmore, G. 1995, \aj, 110, 2771 (WG)
\reference{y1} Yoss, K. M., Neese, C. L., \& Hartkopf, W. I. 1987, \aj, 94, 1600
\reference{y2} Yoshii, Y., Ishida, K., \& Stobie, R. S. 1987, \aj, 93, 323
\reference{z1} Zinn, R. 1996, in Formation of the Galactic
Halo, Inside and Out, ASP Conf. Ser. Vol. 92, edited by H. Morrison and
A. Sarajedini (ASP, San Francisco), p. 211
\end{references}
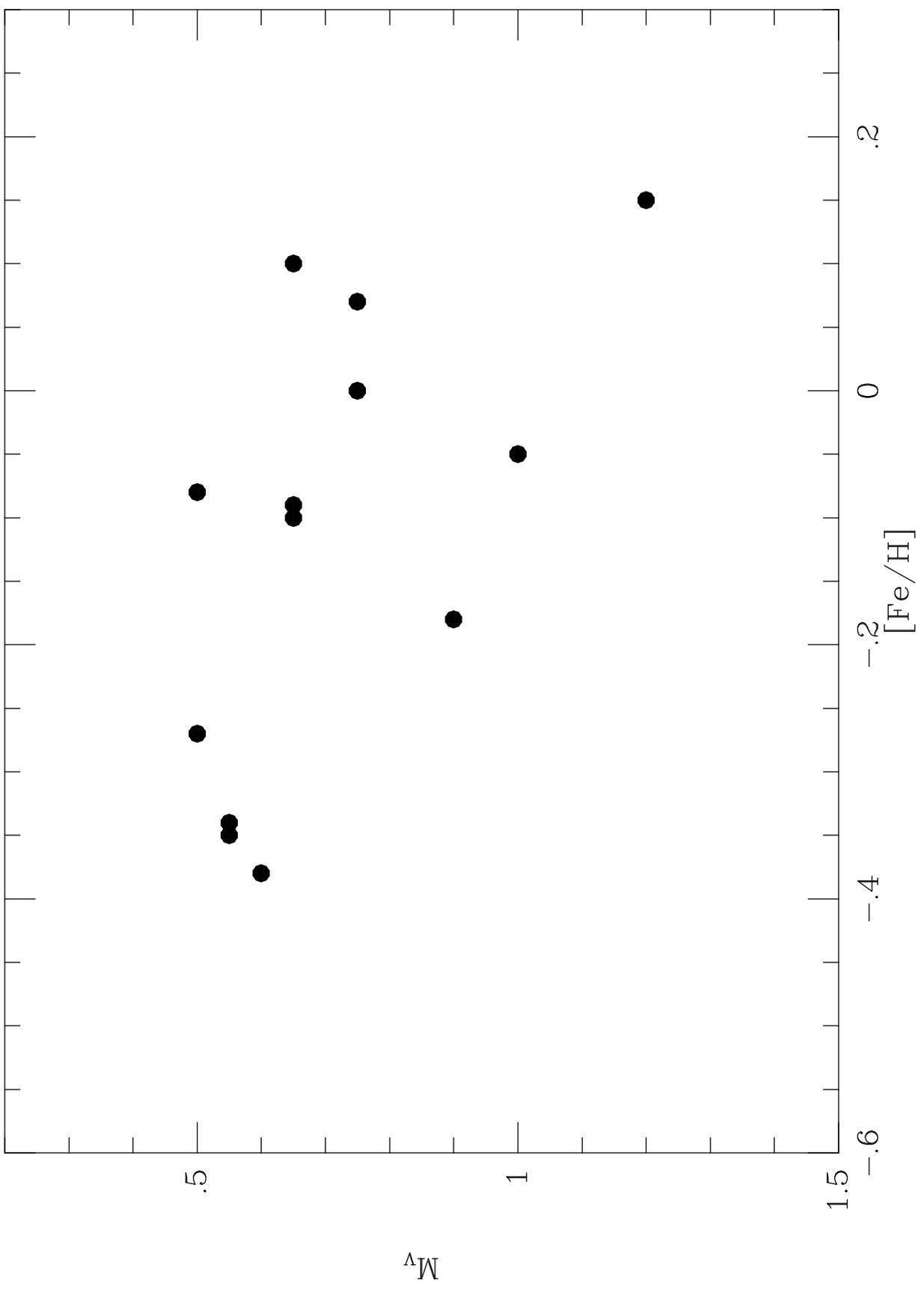
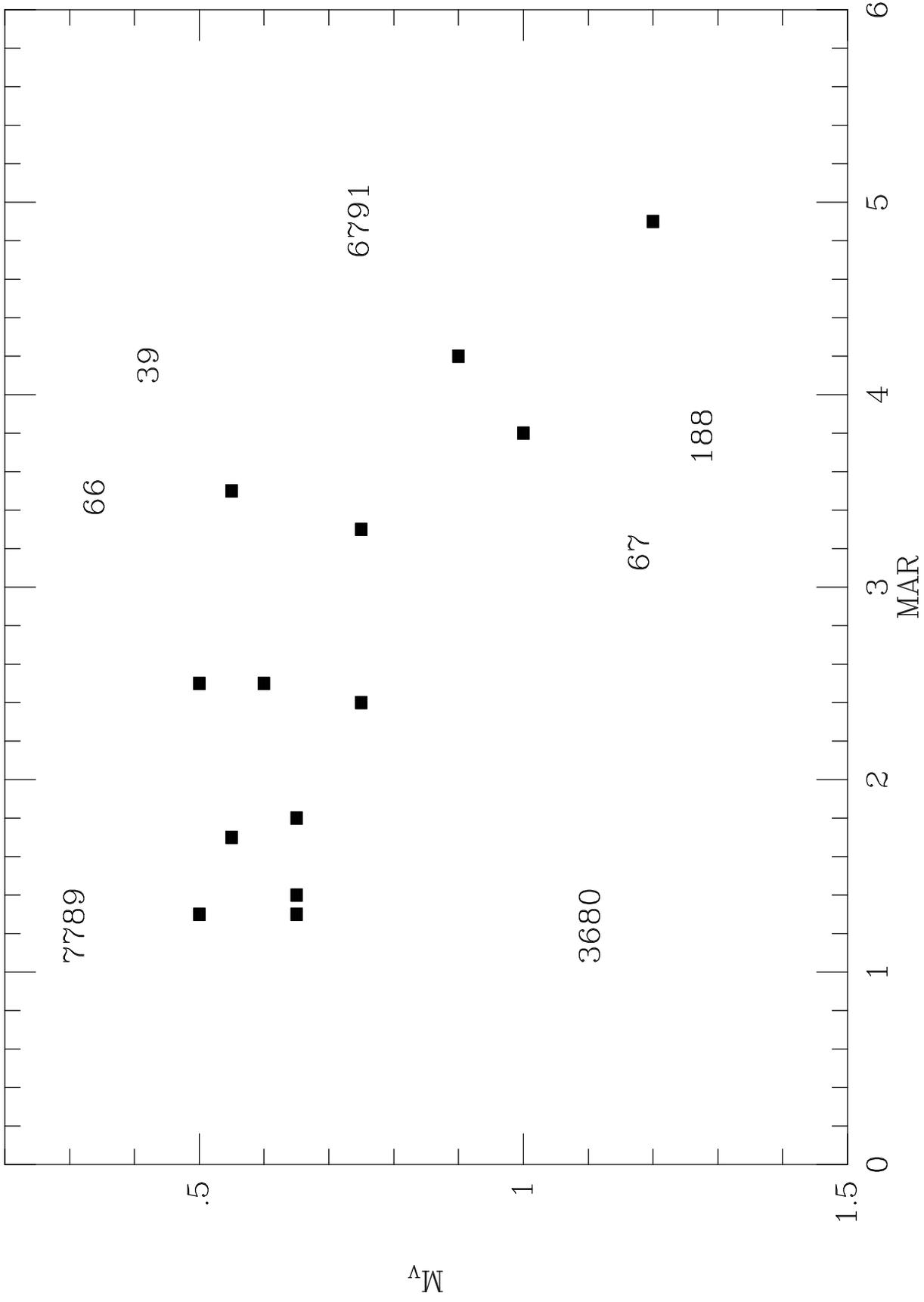
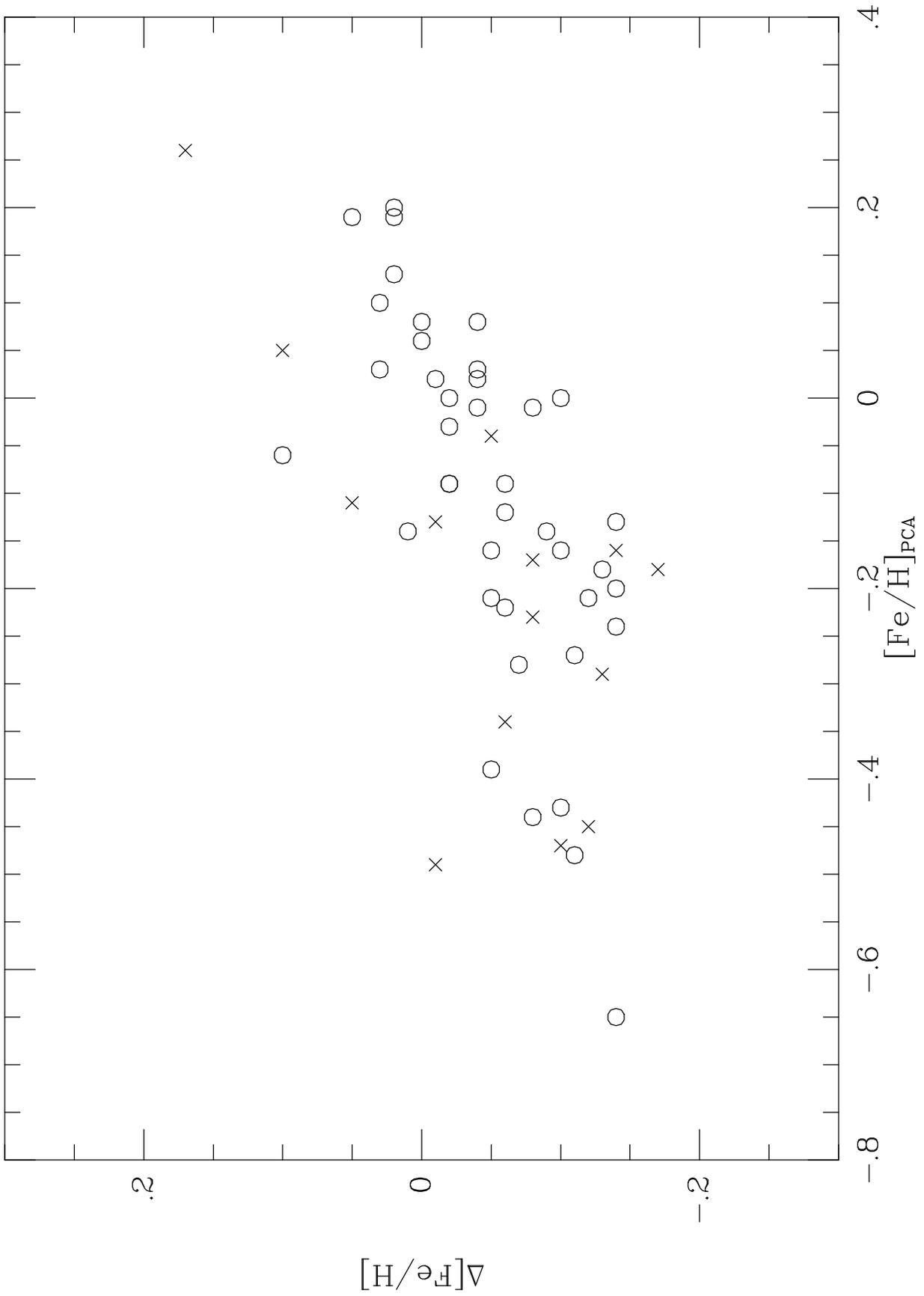
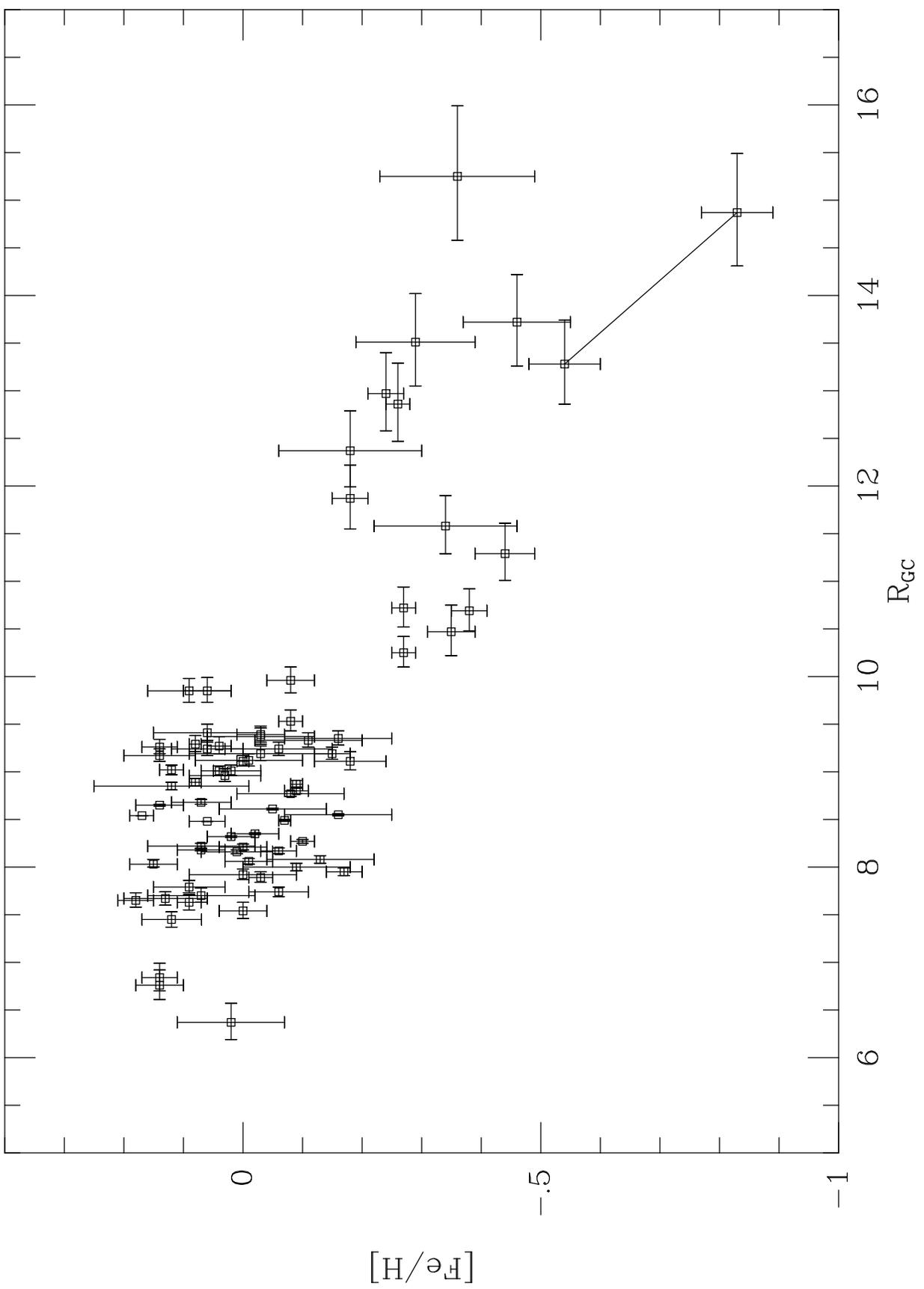
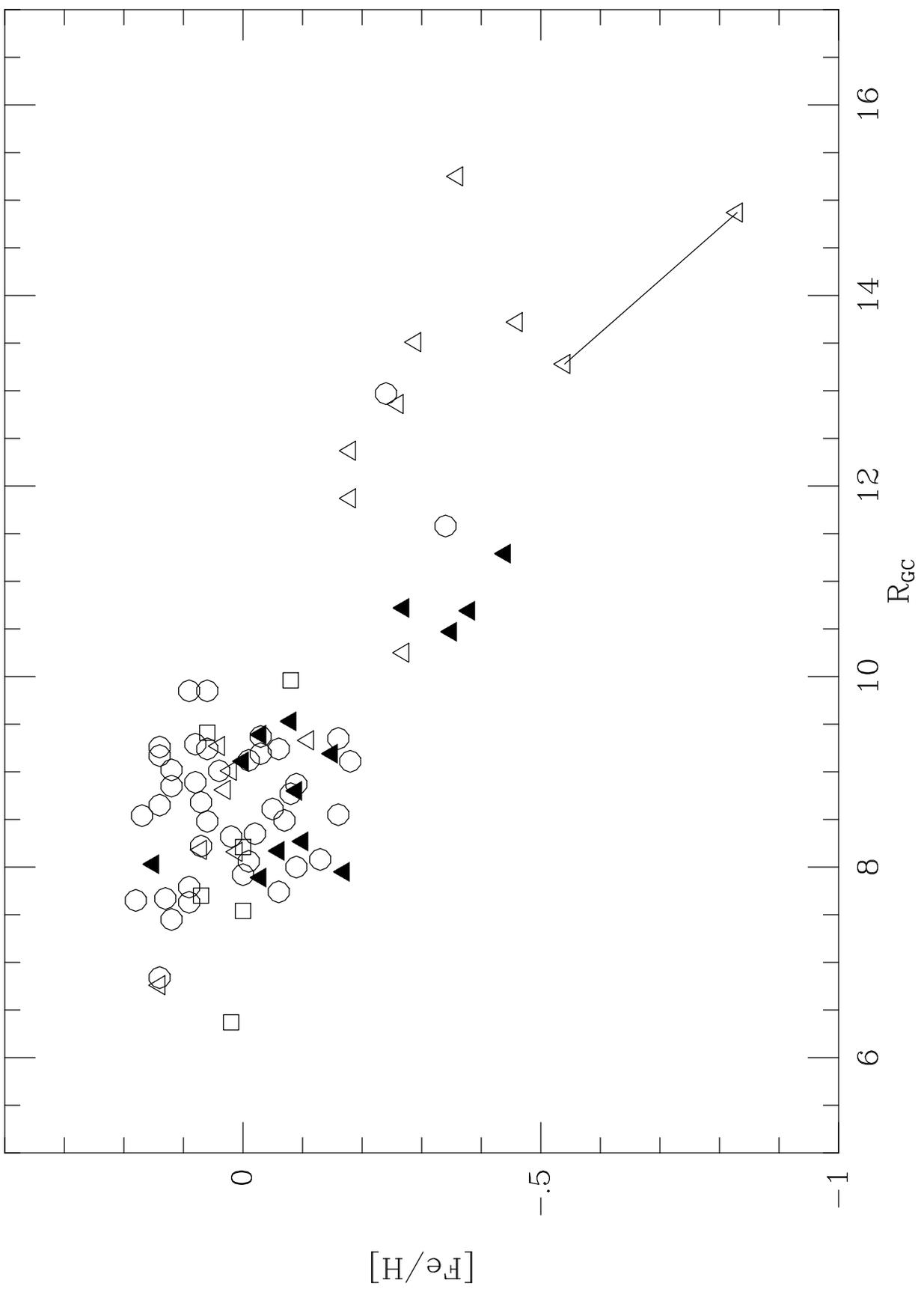
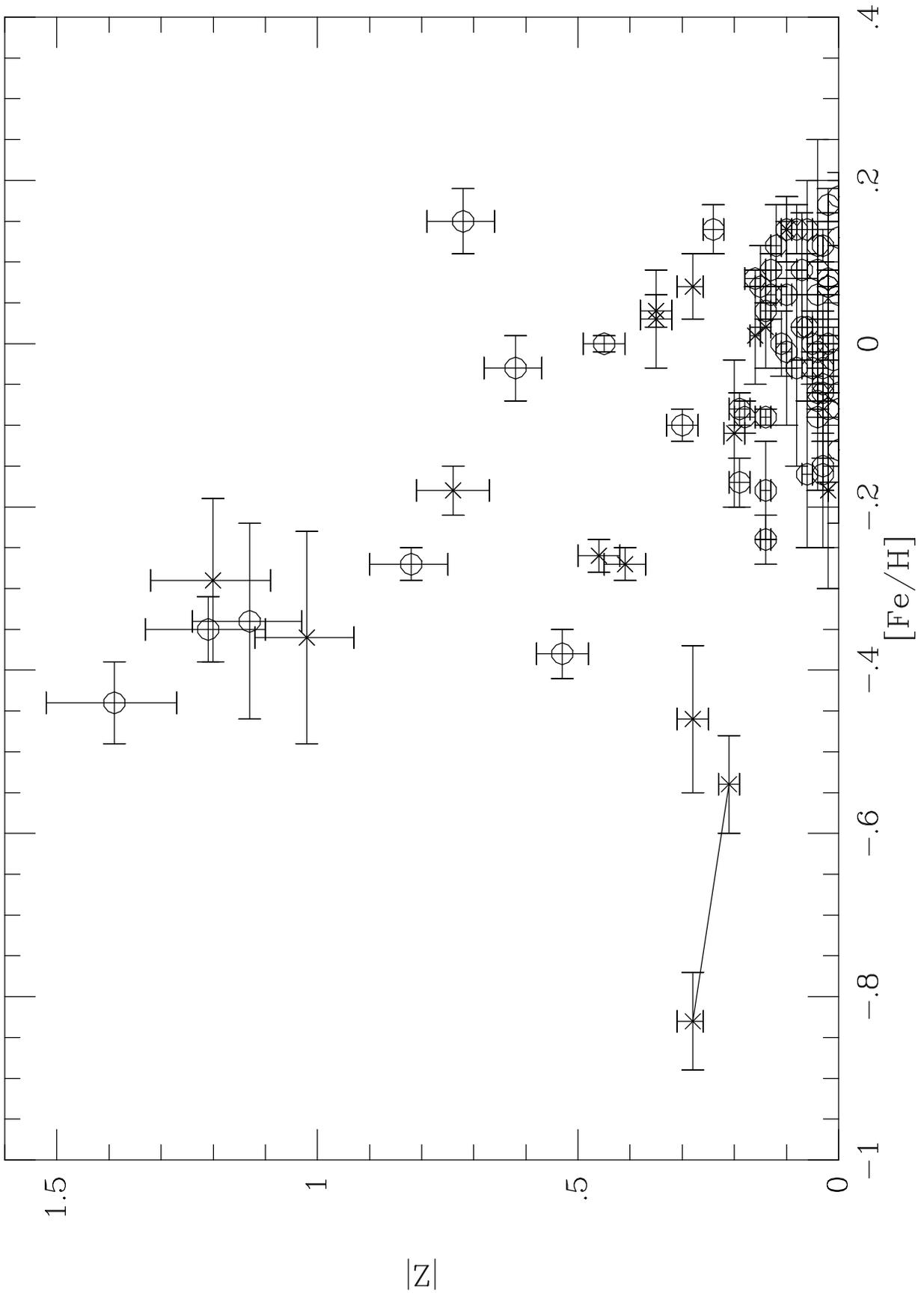
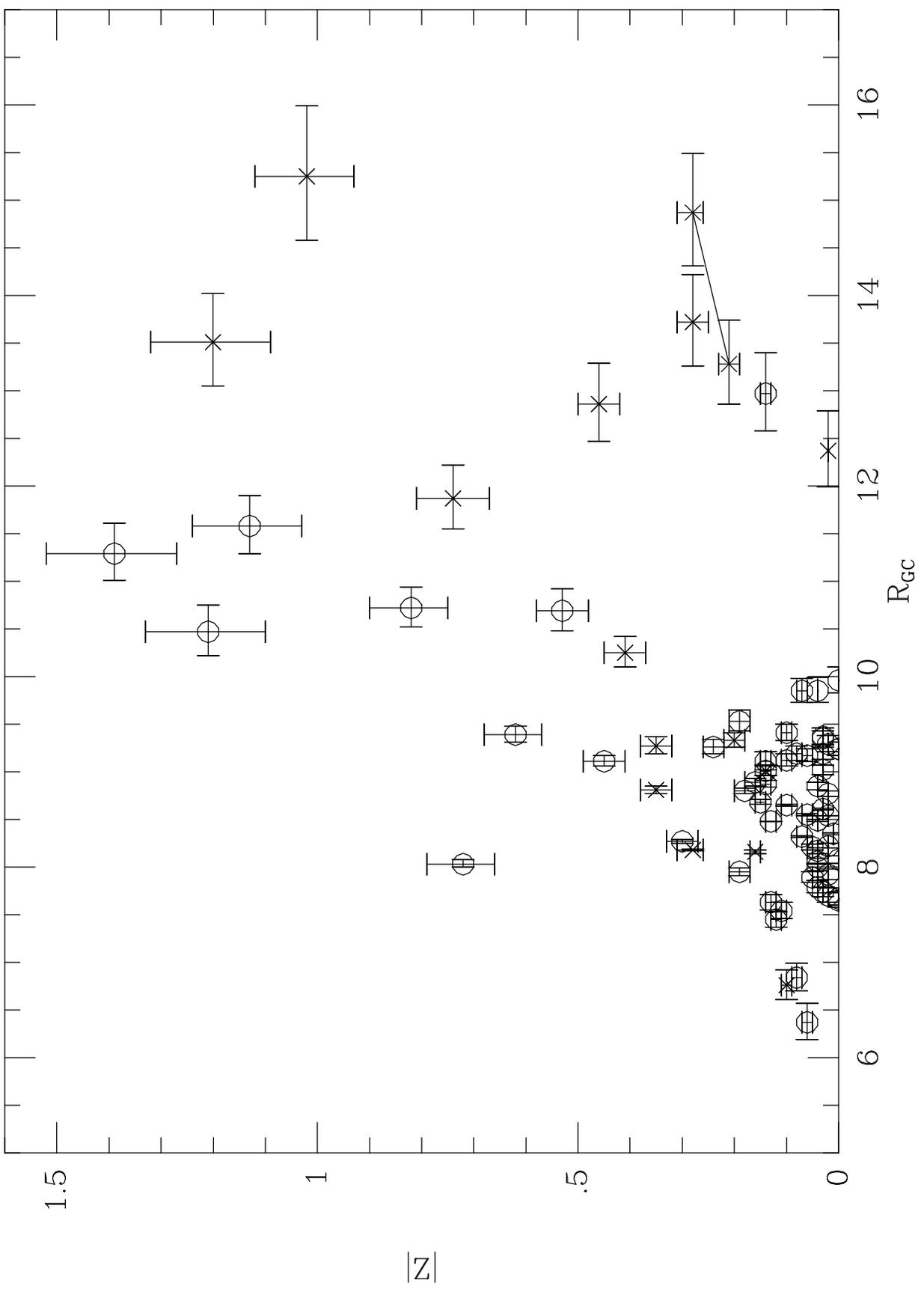
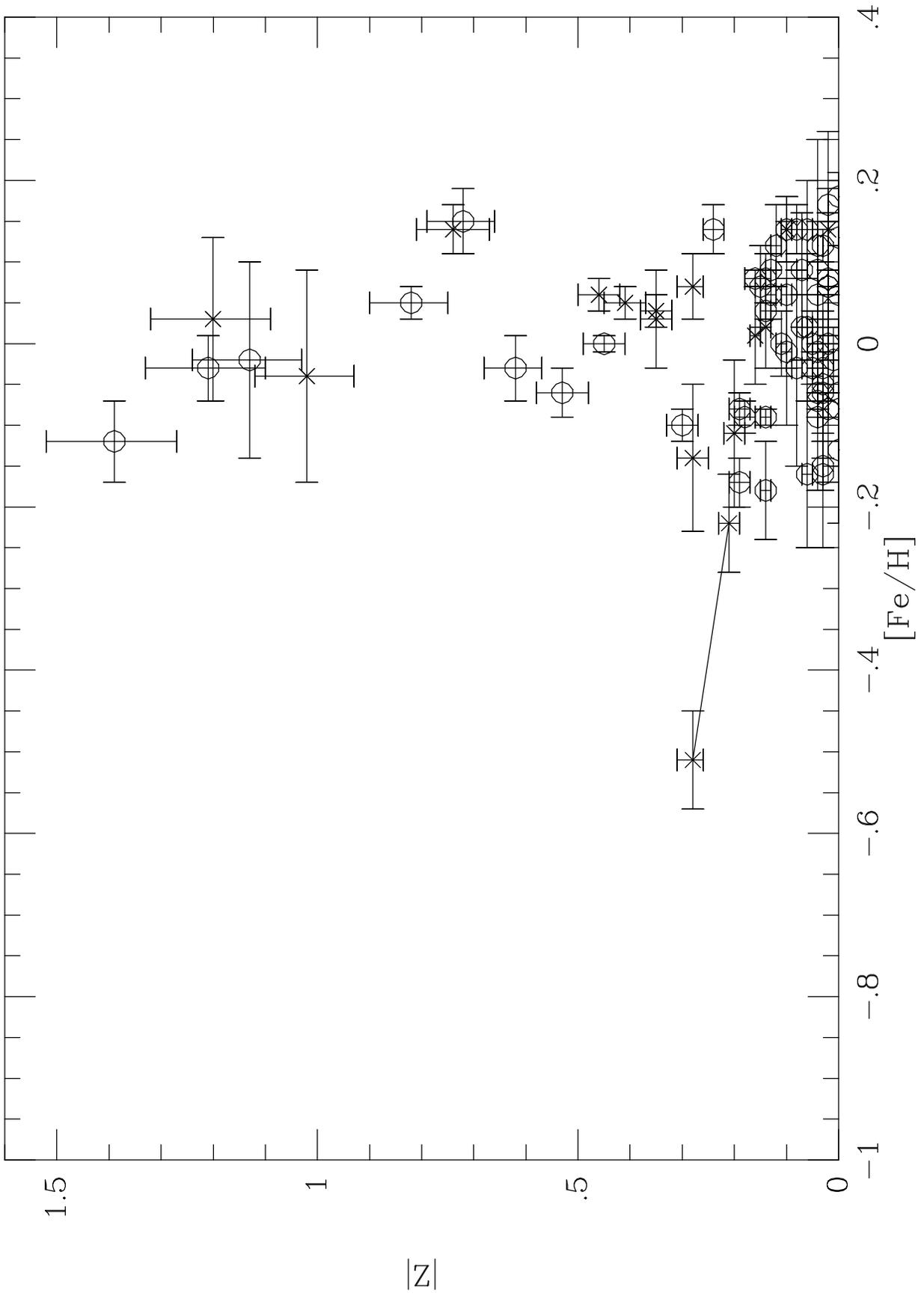
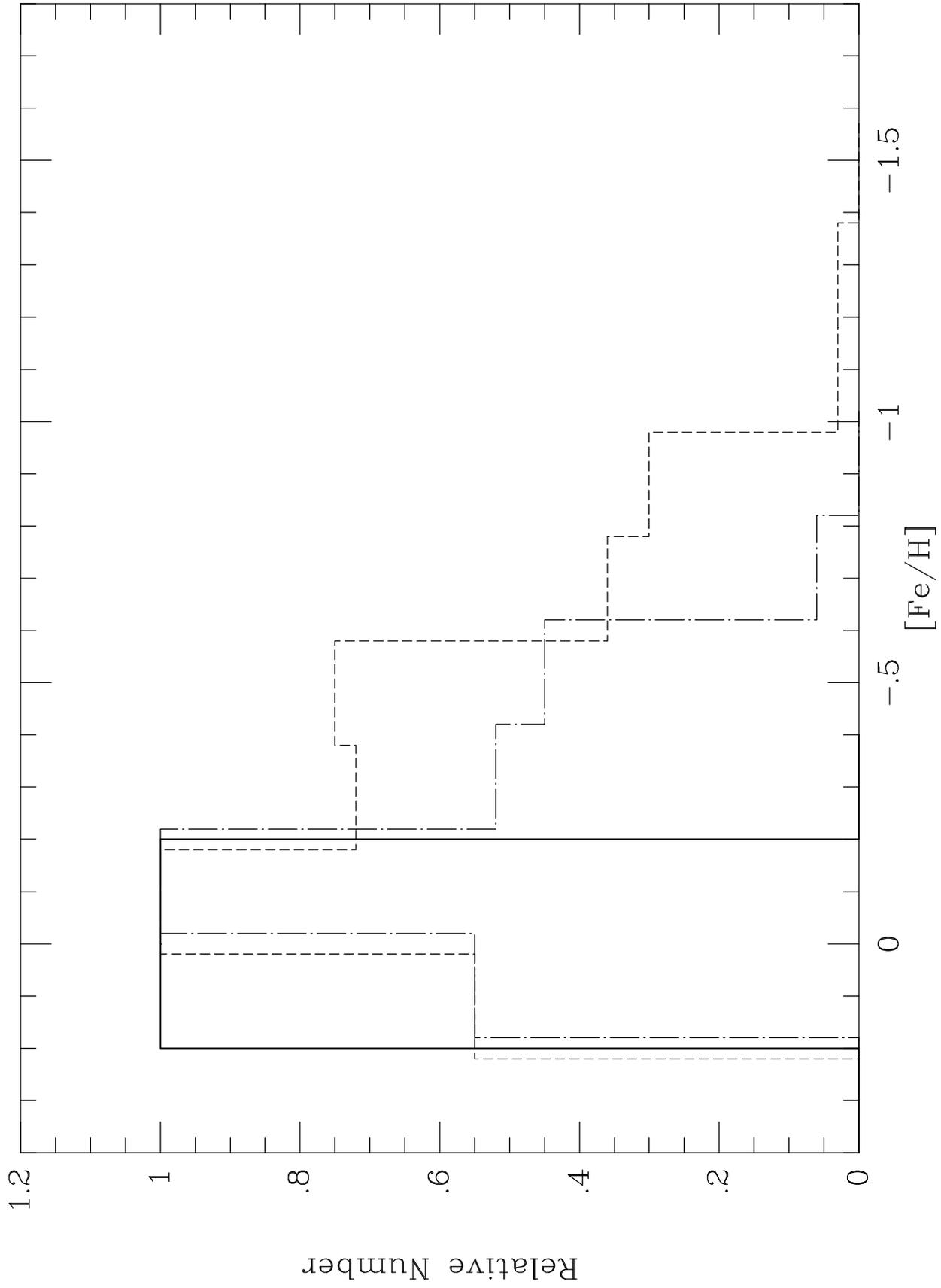
\end{document}